# Multiscale, Techno-economic Evaluation of Isoreticular Series of CALF-20 for Biogas Upgrading using a Pressure/Vacuum Swing Adsorption (PVSA) Process

Changdon Shin[a], Sunghyun Yoon[a], and Yongchul G. Chung*[ab]

[a]*School of Chemical Engineering, Pusan National University, 46241 Busan, South Korea*

[b]*Graduate School of Data Science, Pusan National University, 46241 Busan, South Korea*

E-mail: drygchung@gmail.com

† Electronic Supplementary Information (ESI) available: Adsorption isotherm models, PVSA cycle optimization equations, structural parameters and adsorption enthalpies, single-component adsorption isotherms, isotherm fitting results, isotherm parameters

## Abstract

Cyclic swing adsorption processes, such as pressure/vacuum swing adsorption (PVSA), is a promising technology for upgrading biogas by separating carbon dioxide ($CO_2$) from methane ($CH_4$). The rational design of adsorbent materials with tailored properties is important for deployment of high-performance PVSA technology. Metal–organic frameworks (MOFs), particularly the CALF-20 isoreticular series, have attracted interest due to their high $CO_2$ selectivity, thermal and water stability. In this study, we report a multiscale assessment of CALF-20 and its isoreticular five derivatives by integrating molecular simulations with PVSA process optimization and techno-economic analysis. Structural and adsorption characteristics were calculated and employed to assess how each material performs in terms of energy efficiency and cost. The analysis reveal distinct differences in cost performance among the CALF-20 series, with CALF-20 showing the most favorable economics with >97% purity $CH_4$ production cost at \$4.31 per kg of $CH_4$ and energy consumption of 9.35 kWh per kg of $CH_4$. This study demonstrates that the integrated molecular-process optimization framework can effectively guide the search for adsorbent materials for biogas upgrading.

## Introduction

With the growing importance of greenhouse gas reduction due to global warming, the capture and utilization of major greenhouse gases such as carbon dioxide ($CO_2$) and methane ($CH_4$) have emerged as critical research topics in the energy and environmental sectors. Methane, in particular, has a global warming potential ($GWP_{100}$) approximately 40 times higher than that of carbon dioxide, making the effective management of its emission sources essential[1]. The major sources of methane emission into the atmosphere are from organic wastes, such as sewage sludge, food waste, and livestock manure. As a sustainable solution to the issue, biogas technology, which converts waste gases into high-value energy, has gained increasing attention. Modern biogas plants utilize sealed anaerobic digesters to process waste and capture the resulting gas mixture efficiently, thereby minimizing the unintended atmospheric release of methane. The biogas typically composed of 50–60% $CH_4$ and 35–45% $CO_2$, which can be purified to high-purity methane, offering one of the most practical alternatives to fossil-derived natural gas[2]. Accordingly, advances in biogas upgrading not only promote waste-to-energy valorization but also contribute directly to greenhouse gas reduction and the decarbonization of energy systems[3]. The steady global increase in biogas production and utilization in recent years further underscores the strategic importance of this technology[4].

Representative biogas separation technologies include absorption, membrane separation, cryogenic separation, and pressure/vacuum swing adsorption (PVSA), each differing in separation principles, energy consumption, and operational stability[5-10]. Among these technologies, PVSA has attracted attention as an economically viable option due to its relatively low energy consumption, high product purity, immunity to corrosion, and the absence of chemical solvents or water usage[9, 11]. PVSA is also highly adaptable to variable biogas compositions[12], enabling optimized separation performance through the selection of operating conditions and adsorbent materials[13, 14]. Over the past years, advancements in cycle design[15], the implementation of high-performance adsorbents[16-18], and the incorporation of various optimization strategies[19, 20] have gradually enhanced the competitiveness of PVSA processes over other separation technology. The performance of PVSA processes is largely determined by the structural characteristics of the adsorbent material and its dynamic characteristics during adsorption and desorption cycles[21, 22]. In this context, multiscale modeling, which bridges molecular-level atomistic simulation with process-level modeling, has become a key strategy for evaluating adsorbent materials performance under process conditions[23-28].

Given the central role of adsorbent materials in PVSA cycle, recent research has focused extensively on MOFs as promising candidates for gas separation. In this context, significant efforts have been devoted to developing high-performance MOF adsorbents and applying them to PSA processes for selectively removing $CO_2$ and $N_2$ from methane-based gas mixtures, thereby enhancing $CH_4$ purity. To this end, various studies have employed isotherm measurements, breakthrough experiments, and simulation-based evaluations of separation performance. Additionally, some works have adopted multiscale approaches that bridge molecular-level simulations with process-level modeling. For example, Singh et al. investigated the $CO_2/N_2$ and $CH_4/N_2$ selectivity of a $Fe_4O_2$-based MOF (IISERP-MOF30) using GCMC and MD simulations, and proposed a two-step PSA process for natural gas upgrading[29]. Karimi et al. experimentally quantified the $CO_2/CH_4$ and $CO_2/N_2$ adsorption performance of MIL-160(Al) using breakthrough experiments and response surface methodology (RSM)[17], and further applied its pelletized form to a PSA setup, achieving 99% $CH_4$ purity and 63% recovery[30]. Abd et al. optimized the PSA performance of UiO-66 under various operating conditions through experimental and dynamic modeling, obtaining 99.99% $CH_4$ purity, 99.99% recovery, and a productivity of 8.57 mol/kg/h[31]. More recently, Rogacka et al. performed a systematic numerical screening of MOFs for biogas purification under humid conditions, demonstrating that the presence of water can significantly affect adsorption capacity, regenerability, and selectivity[32]. These studies demonstrate the importance of quantitatively linking the structural characteristics of MOF adsorbents with process-level parameters, highlighting the critical role of material selection in designing high-performance PSA systems for methane purification. In addition, techno-economic assessments of biogas upgrading, including PVSA-based processes[33] and market-level evaluations[34], have also been reported, underscoring that process performance needs to be coupled with cost and scalability considerations.

Recently, the isoreticular series of CALF-20 has been computationally proposed in the literature as promising $CO_2$ capture materials under post-combustion ($CO_2/N_2$) conditions, owing to its high $CO_2$ selectivity in the presence of water[35, 36]. CALF-20 is a highly thermal and chemically stable zinc triazolate MOF made of 1,2,4-triazolate-bridged Zin(II) layers pillared by the oxalate ligand. This adsorbent material is the current benchmark MOF sorbent for $CO_2$ capture from cement flue gas conditions and has been scaled up and deployed in a pilot plant[37]. Inspired by this material, Gopalsamy et al. computationally constructed isoreticular series of CALF-20 derivatives[36]. The oxalate ligand of the parent CALF-20 was substituted by alternative small linkers, including squarate (Squ), fumarate (Fum), benzenedicarboxlate (Bdc), thieno[3,2-b]thiophene-2,5-dicarboxlate (Ttdc), and cubanedicarboxlate (Cub). Based on the GCMC simulations, they found that SquCALF-20 is the best candidate combining high $CO_2$ uptake at 0.15 bar (3.6 mmol/g) and very high $CO_2/N_2$ selectivity (500), exceeding the performance of pristine CALF-20. However, the potential for biogas upgrading for the isoreticular series CALF-20 materials under process conditions has not yet been evaluated.

We evaluate the performance of isoreticular CALF-20 adsorbent materials by integrating atomistic molecular modeling with PVSA process simulation and multi-objective optimization using the TSEMO algorithm. The geometric structures of the CALF-20 derivatives were first optimized using the MACE machine learning potential, and the GCMC simulations were carried out to generate single-component isotherms. These isotherms were subsequently fitted to the dual-site Langmuir (DSL) equation, and the PVSA simulations were performed to evaluate the process-level performance.

# Methods

## 1. Crystal structures

A pristine CALF-20 and a series of its ligand-substituted derivatives proposed by Gopalsamy et al.[36] were considered. The crystal structure of CALF-20 was obtained from the Cambridge Crystallographic Data Centre (CCDC) (REFCODEL: TASYAR)[37]. A total of five ligand-substituted derivatives were constructed by replacing the oxalate linker in CALF-20 with squarate (Squ), fumarate (Fum), benzenedicarboxylate (Bdc), thieno[3,2-*b*]thiophene-2,5-dicarboxylate (Ttdc), and cubanedicarboxylate (Cub), using PoreMatMod.jl[38]. CALF-20 and its ligand-substituted derivatives were geometrically optimized using a two-step cell optimization protocol. Firstly, initial cell optimizations were performed using the Forcite module in Materials Studio[39] using the Universal Force Field (UFF)[40] to obtain physically reasonable initial crystal structures. Ligand substitutions were then carried out on these pre-optimized structures using PoreMatMod.jl to ensure consistent geometry generation. After each substitution, the resulting structures were re-optimized to relax the framework geometry before subsequent simulations. These structures were further refined to higher accuracy level theory through final cell optimization using the atomic simulation environment (ASE)[41] package with the MACE-MP-0 model[42]. The BFGS algorithm was applied with a force convergence threshold of 0.001 eV/Å during the final optimization. The geometric properties, including pore volume, pore limiting diameter (PLD), and largest cavity diameter (LCD), were calculated using the Zeo++ software package[43]. All relevant input files for the structure generation are available at https://github.com/Chung-Research-Group/reproducible-workflows/2025-Biogas.

## 2. Molecular simulation

The single-component adsorption isotherms of $CO_2$ and $CH_4$ in an isoreticular series of CALF-20 materials were obtained by carrying out GCMC simulations. Non-bonded interactions between adsorbate molecules and framework atoms were modeled with a Lennard–Jones (12-6) potential plus

Coulomb terms, truncated at a 14.0 Å cutoff and with tail corrections. LJ parameters for the framework atoms of the adsorbent were taken from the DREIDING force field[44], while those for the adsorbates were obtained from the TraPPE force field[45, 46]. Interactions between different atom types were approximated using the Lorentz–Berthelot mixing rules. $CO_2$ was modeled as a rigid three-site molecule consisting of one carbon and two oxygen interaction sites, whereas $CH_4$ was represented as a single-site united atom model. To satisfy the minimum image convention and prevent self-interactions between atoms, periodic boundary conditions were imposed by replicating the unit cell along the x, y, and z directions such that the shortest box length exceeded twice the vdW cutoff distance (28.0 Å). Partial atomic charges for the framework were assigned using the PACMAN DDEC06 model[47]. Simulations were carried out over a pressure range of 1 to 500,000 Pa at three different temperatures: 273 K, 298 K, and 323 K. Each simulation ran for 20,000 cycles. The first 10,000 cycles were discarded to allow for the system to reach equilibrium, and the remaining 10,000 cycles were averaged to calculate statistically meaningful uptake values. Each cycle included a minimum of 20 Monte Carlo (MC) moves, or number of adsorbate molecules inside the simulation box, whichever is greater. These MC moves consisted of translation, rotation, reinsertion, and swap moves, each performed with equal probability. Additionally, the isosteric heats of adsorption for $CO_2$ and $CH_4$ in six isoreticular series of CALF-20 materials were calculated using Widom particle insertion simulations[48]. For each gas species, 20,000 Monte Carlo cycles were performed to determine the heats of adsorption. All GCMC and Monte Carlo simulations were conducted using the RASPA 2.0 simulation package[49]. All simulations were carried out with a rigid framework assumption where the positions of framework atoms do not change during the course of the simulation.

### 3. Adsorption isotherm model

The adsorption isotherms of $CO_2$ and $CH_4$ was fitted to an analytical model equation based on the assumption that two types of adsorption sites exist on the adsorbent surface, each exhibiting different adsorption strengths. Single-component adsorption isotherms were fitted using the dual-site Langmuir (DSL) model, which distinguishes between strong and weak binding sites[50, 51]. The DSL model parameters were obtained by fitting the single-component $CO_2$ and $CH_4$ isotherms, with site capacities constrained to ensure thermodynamic consistency. For mixture isotherm predictions, both the extended dual-site Langmuir (EDSL) model and the Ideal Adsorbed Solution Theory (IAST)[52] were applied. The DSL and EDSL model equations are provided in Section S1 of the Supplementary Information, along with the IAST formulation. To validate the mixture predictions from these models, $CO_2$/$CH_4$ binary adsorption isotherms were also computed by GCMC simulations. Selectivity ($\alpha$) is defined as the ratio of the molar loading of $CO_2$ to $CH_4$, normalized by their respective gas-phase mole fractions at a given pressure:

$$\alpha = \frac{(q_A/q_B)}{(y_A/y_B)} \tag{1}$$

The working capacity (WC) is defined as the difference in gas uptake between adsorption and desorption pressures, reflecting the usable adsorption capacity within a PVSA cycle:

$$WC_A = q_{ads,A} - q_{des,A} \tag{2}$$

We set adsorption and desorption pressures to 1 bar and 0.1 bar, respectively, to calculate the working capacity used throughout this work.

## 4. Pressure/vacuum swing adsorption (PVSA) cycle

Based on the mixture prediction methods above, we performed a modified 5-step Skarstrom cycle as the PVSA simulation model, consisting of pressurization, adsorption, heavy reflux, counter-current depressurization, and light reflux (Scheme S1). This configuration was selected because it effectively balances $CH_4$ recovery and $CO_2$ removal under PSA-relevant conditions. The feed gas was assumed to be a binary mixture of $CH_4$ (55%) and $CO_2$ (45%), representing a typical composition from biogas sources such as wastewater sludge and livestock manure digestion. During the cycle, $CO_2$ is preferentially adsorbed, and the product stream produces high purity $CH_4$. A detailed schematic and stepwise description of the cycle are provided in the Supplementary Information Section S2.

## 5. PVSA modeling

The adsorption column in this PVSA model follows the same assumptions as the MATLAB model developed by Leperi et al., which are as follows[50, 53]:

1) The gas phase follows the ideal gas law.
2) There is axially dispersed plug flow in the column
3) The gas and solid phases are in thermal equilibrium.
4) No radial gradients in concentration, temperature, or pressure are considered.
5) The solid-phase mass transfer rate is represented using the linear driving force (LDF) model.
6) The pressure drops along the column is calculated using Ergun's equation.
7) No heat transfer occurs across the column wall.
8) The void fraction and particle size remain constant along the column.

We used a one-dimensional, non-isothermal, and non-isobaric dynamic column to model the adsorption column. The transport phenomena within the column were described by a set of coupled fundamental equations, including component mass balance, total mass balance, energy balance, and momentum balance. Equilibrium mixture adsorption behavior was described by the dual-site Langmuir isotherm or IAST as described above. The energy balance accounts for convection, conduction, heat generated by adsorption, and heat losses to the surroundings. To improve computational efficiency, the resulting set of coupled partial differential equations (PDEs) was nondimensionalized using appropriate scaling factors. The spatial domain was discretized using the finite volume method (FVM) with a weighted essentially non-oscillatory (WENO) scheme[54]. This led to a system of time-dependent ordinary differential equations (ODEs), which were numerically integrated using the stiff ODE solver ode15s available in MATLAB. The dynamic model comprises a set of dimensionless governing equations describing overall mass conservation, component mass balances, adsorption kinetics, pressure drop, and energy balances for both the gas phase and the column wall. These equations account for axial dispersion, convection, adsorption–desorption dynamics, and heat transfer phenomena such as conduction, convection, heats of adsorption, and external heat transfer. The full set of nondimensional model equations and parameter definitions are provided in Section S2 of the Supplementary Information.

To simulate the full adsorption cycle, a commonly adopted uni-bed modeling approach was employed. In this method, a single adsorption column sequentially undergoes all steps of the cycle in a time-dependent manner. This approach enables efficient evaluation of the overall cycle performance while significantly reducing computational cost, as it avoids modeling multiple columns in parallel. The simulation proceeds through repeated cycles until the system reaches cyclic steady state (CSS). CSS was defined as the condition where the relative differences between the dimensionless variables (mole fraction, pressure, temperature, and molar loading) at the end of the final step and the beginning of the first step in a cycle were all within 1%. Additionally, the total mass balance error, defined as the difference between the total mass input and output over one cycle, was required to be less than 0.01. The maximum number of cycle iterations was limited to 150. Simulations that did not meet the CSS criteria within this limit were considered non-converged and excluded from further analysis.

## 6. Process technical performance indicators

To evaluate the performance of adsorption-based separation processes, it is essential to quantitatively define the product composition, the extent of recovery relative to the feed, and the energy consumption required for operation. In this study, $CO_2$ purity, $CO_2$ recovery, and energy consumption were employed as the fundamental technical performance indicators. These metrics not only represent the fundamental performance of the separation process but also impact the techno-economic feasibility of the process. The evaluation of these performance indicators requires both process-level operating conditions and adsorbent-specific physical properties. Table 1 summarizes the process parameters, including adsorption pressure, feed velocity, and reflux ratios, which were optimized during the PVSA simulation. Table 2 provides the intrinsic physical properties of the adsorbent materials, such as crystal density and specific heat capacity, calculated from the MACE-optimized structures, used as inputs to the process simulation. The heat capacities of the MOF adsorbents were predicted using XGBoost-based machine learning models by Moosavi et al.[55, 56], and these values were applied as constant inputs across all PVSA simulations to capture the thermal and mass-transfer behavior of each adsorbent material. The detailed definitions for $CH_4$ purity, $CH_4$ recovery, and energy consumption are provided in the Supplementary Information (Section S3).

**Table 1.** Parameters and decision variables with lower and upper bounds used in simulations and optimizations of the PVSA cycle.

| **Parameter** | **Unit** | **Value** | **Type** |
|---|---|---|---|
| Column void fraction | - | 0.37 | Constant |
| Viscosity of gas | Pa·s | $1.28 \times 10^{-5}$ | Constant |
| Specific heat of gas | J/mol/K | 36.7 | Constant |
| Radius of the pellets | m | $1 \times 10^{-3}$ | Constant |
| Molecular diffusivity | $m^2/s$ | $1.30 \times 10^{-5}$ | Constant |
| Thermal conduction in gas phase | W/m/k | 0.09 | Constant |
| Mass transfer coefficient for $CH_4$ | 1/s | 0.33 | Constant |
| Mass transfer coefficient for $CO_2$ | 1/s | 0.16 | Constant |
| Feed temperature | K | 298.15 | Constant |
| Feed composition ($CO_2$:$CH_4$) | - | 45:55 | Constant |
| Column length | m | [2, 9] | Variable |
| Adsorption pressure | bar | [1, 10] | Variable |
| Desorption pressure | bar | [0.1, 0.5] | Variable |
| Pressurization time | S | [10, 100] | Variable |
| Depressurization time | s | [10, 100] | Variable |
| Feed time | s | [10, 1000] | Variable |
| Feed velocity | m/s | [0.1, 2] | Variable |
| Light reflux ratio | - | [0.01, 0.99] | Variable |
| Heavy reflux ratio | - | [0, 1] | Variable |

**Table 2.** Physical parameters of adsorbent materials used in simulations of the PVSA cycle.

| Materials | Crystal Density (kg/m$^3$) | Heat Capacity (J/kg/K) |
|---|---|---|
| CALF-20 | 1515.86 | 727.6 |
| SquCALF-20 | 1380.50 | 733.4 |
| FumCALF-20 | 1176.39 | 744.7 |
| BdcCALF-20 | 1113.45 | 775.1 |
| CubCALF-20 | 1197.47 | 795.9 |
| TtdcCALF-20 | 1107.91 | 776.6 |

## 7. Process economic performance indicators

We conducted a techno-economic analysis of the adsorption process, which involves consideration of the scaling up of the adsorption column using parallel unit train operations.

### 7.1. Unit train design

The procedure of Khurana and Farooq (2017)[51] was adopted to design the unit train for continuous PVSA operation. In a cyclic process, a single adsorption column alternates between adsorption and regeneration and therefore cannot continuously process the feed. To guarantee uninterrupted operation, multiple columns must be arranged in parallel and operated in different cycle steps. The sizing of auxiliary equipment was considered first. Since the feed stream is directed to one column at a time, only a single feed compressor is required. Similarly, as adsorption and heavy reflux occur simultaneously, one heavy-reflux compressor is sufficient per train. In contrast, the counter-current depressurization and light reflux steps can overlap across columns, requiring additional vacuum pumps. The detailed equations used to calculate the minimum numbers of vacuum pumps, columns per train, and idle step duration are provided in the Supplementary Information (Section S5.1.).

### 7.2. Parallel unit trains

When the required feed throughput exceeds the handling capacity of a single unit train, additional trains are deployed in parallel. This parallelization strategy enables large-scale PVSA operation without disrupting the cycle sequence across columns. The average molar flow of methane into a single train was calculated from the pressurization and adsorption intervals, and the number of parallel trains required to process the total $CO_2/CH_4$ feed was determined accordingly. The detailed mathematical expressions for these calculations are also provided in the Supplementary Information (Section S5.2.).

### 7.3. Cost model

The capital cost (CAPEX) and operating cost (OPEX) of the PVSA system were estimated following the methodology and correlations reported by Turton et al.[57] and implemented within the techno-economic framework proposed by Yoon et al.[58], with all estimates expressed in US dollars. Major cost components included adsorption columns, vacuum pumps, and compressors, with bare module costs (BMCs) calculated using standard purchase cost correlations. CAPEX was obtained by aggregating equipment costs, contingency, fees, and site-related costs, and subsequently annualized using the equivalent annual cost (EAC) approach. OPEX was estimated as the sum of electricity consumption, adsorbent replacement, labor, maintenance, supervisory expenses, operating supplies, administrative overhead, and plant overhead. Among these, electricity costs associated with vacuum pumps and compressors were the most significant contributor. Finally, the methane production cost was calculated based on the total annualized cost (TAC), defined as the sum of EAC and OPEX, normalized by the recovered methane throughput. The detailed correlations and calculation equations used for CAPEX, OPEX, and $CH_4$ production cost are provided in the Supplementary Information (Section S5.3.). The economic parameters applied in the estimation of both EAC and OPEX are summarized in Table 3.

**Table 3.** Key parameters for techno-economic evaluation.

| Parameter | Unit | Value |
|---|---|---|
| Discount rate[59], $d$ | – | 0.08 |
| Economic lifetime[59], $t$ | yr | 25 |
| Electricity unit cost[57], $\boldsymbol{C_{elec}}$ | \$/kWh | 0.07 |
| Adsorbent cost[19], $\boldsymbol{C_{ads}}$ | \$/kg | 10 |
| Chemical engineering plant cost index, CEPCI | | |
| 2024 | – | 798.8 |
| 2001 | – | 397 |

## 8. Cost optimization

The optimal design and operating conditions that minimize the $CH_4$ production cost in the PVSA system were identified through $CH_4$ production cost optimization. Among the obtained solutions, only those with $CH_4$ purity ≥ 97% were considered, and the lowest-cost case within this feasible set was identified as the optimum. Eight operating variables, namely the adsorption pressure ($P_H$), feed velocity ($\boldsymbol{v}_F$), desorption pressure ($P_L$), light reflux ratio ($\alpha_{LR}$), heavy reflux ratio ($\beta_{HR}$), adsorption time ($t_{ads}$), counter-current depressurization time ($t_{depres}$), and pressurization time ($t_{pres}$), together with one design variable, the column length ($L$), were chosen as decision variables. The column length-to-diameter ratio ($L/D$) was fixed at 3[59], and the bounds for the decision variables are summarized in Table 2.

We employed TSEMO[19] to optimize the process. The implementation of TSEMO was adapted from the open-source MATLAB code[60]. The initial dataset size was set to 150, and the algorithm was configured to run for 200 consecutive iterations. TSEMO enables efficient exploration of high-dimensional objective spaces with a minimal number of simulations, allowing for performance evaluation and comparison of adsorbents based on the resulting Pareto fronts[61]. To account for accurate mixture adsorption from $CH_4/CO_2$ mixtures, the optimization framework was based on the IAST model fitted to single component GCMC data.

## Results and discussion

### 1. Structural optimization and pore characteristics

The optimized crystal structures of CALF-20 and its derivatives are presented in Fig. 1 and highlight the pore-level structural differences with the incorporation of different organic linkers. These structural differences influence overall material properties such as pore size, pore volume, and adsorption enthalpy, which leads to variations in adsorption isotherms and ultimately process performance.

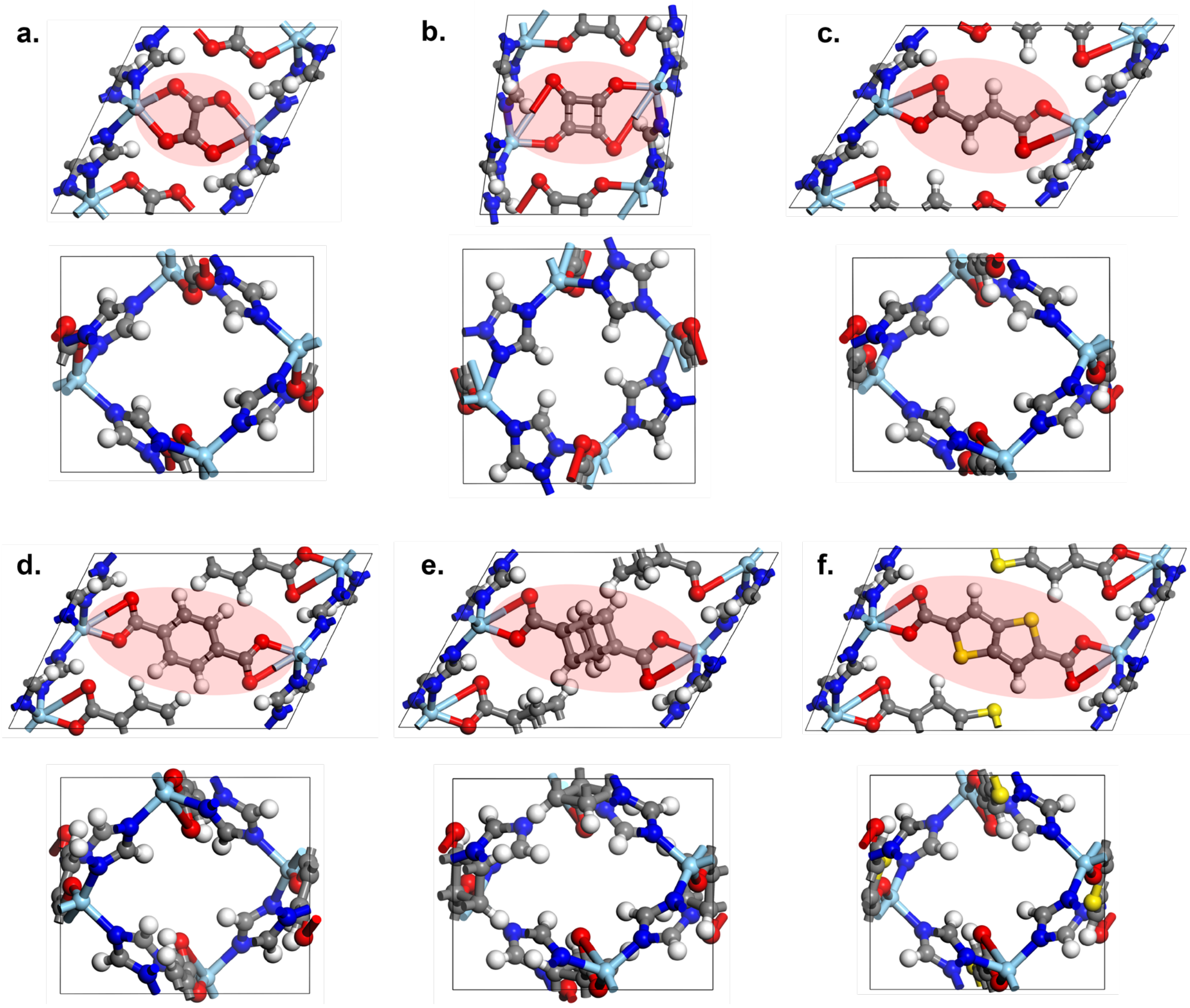

**Fig. 1** Side views (top) and channel views (bottom) of optimized: **a.** CALF-20, **b.** SquCALF-20, **c.** FumCALF-20, **d.** Bdc-CALF-20, **e.** CubCALF-20, and **f.** Ttdc-CALF-20. Atom colors: gray (C), red (O), blue (N), white (H), yellow (S), and light blue (Zn).

Table 4 summarizes the pore characteristics and $CO_2/CH_4$ adsorption enthalpies of each adsorbent material. The pore volume ranges from 0.35 to 0.56 cm³/g, with CALF-20 exhibiting the lowest pore volume and TtdcCALF-20 the highest. Despite having a relatively large LCD of 4.7 Å, SquCALF-20 has a narrow PLD of 2.9 Å, which restricts the accessibility of adsorbate molecules. TtdcCALF-20, with the highest pore volume and LCD (0.56 cm³/g and 5.0 Å, respectively), provides a more open pore characteristic. However, TtdcCALF-20 shows a relatively moderate adsorption affinity, with a $CO_2$ adsorption enthalpy of –28.7 kJ/mol, almost 10 kJ/mol lower than CALF-20 (–36.3 kJ/mol). FumCALF-20 and BdcCALF-20 materials, both feature large pore volumes (0.52 cm³/g and 0.55 cm³/g, respectively), with PLD and LCD values above 3.1 Å. These materials show moderate $CO_2$ adsorption enthalpies (–27.7 and –27.6 kJ/mol, respectively) and relatively low $CH_4$ affinities (–18.9 and –19.8 kJ/mol), which may contribute to improved $CO_2/CH_4$ selectivity. CubCALF-20 shows a relatively high $CO_2$ enthalpy of –31.7 kJ/mol, along with a well-balanced pore structure (PLD of 3.2 Å and LCD of 4.8 Å).

**Table 4.** Structural properties and adsorption enthalpies of adsorbent materials.

| **Materials** | **Pore volume (cm³/g)** | **PLD (Å)** | **LCD (Å)** | **$\Delta H$ (kJ/mol)** | |
|---|---|---|---|---|---|
| | | | | **$CO_2$** | **$CH_4$** |
| CALF-20 | 0.35 | 3.0 | 4.4 | −36.3 | −24.7 |
| SquCALF-20 | 0.40 | 2.9 | 4.7 | −29.4 | −21.6 |
| FumCALF-20 | 0.52 | 3.4 | 5.0 | −27.7 | −18.9 |
| BdcCALF-20 | 0.55 | 3.1 | 4.7 | −27.6 | −19.8 |
| CubCALF-20 | 0.48 | 3.3 | 4.8 | −31.7 | −23.1 |
| TtdcCALF-20 | 0.56 | 3.2 | 5.0 | −28.7 | −20.2 |

## 2. Single-component adsorption isotherms

The adsorption behavior of $CO_2$ and $CH_4$ in CALF-20 and its derivatives were investigated using GCMC simulations. To validate our computational approach, the simulated isotherm for the CALF-20 was compared with available experimental data (Fig. 2).[36] Simulated results for the other materials are provided in the Supplementary Information (Figs. S1 and S2). For $CO_2$, all materials exhibited a steep uptake at low pressures followed by saturation at higher pressures. TtdcCALF-20, BdcCALF-20, and CubCALF-20, which possess structurally open pores and sufficiently large pore openings, showed relatively high $CO_2$ uptake capacities. FumCALF-20 also demonstrated good $CO_2$ adsorption characteristic. In contrast, SquCALF-20 showed the lowest $CO_2$ uptake across the pressure range, likely due to restricted molecular accessibility caused by its narrow pore limiting diameter. For $CH_4$, the overall uptake was lower than that of $CO_2$, yet distinct differences were observed among the MOFs. TtdcCALF-20, BdcCALF-20, FumCALF-20, and CubCALF-20 exhibited relatively high $CH_4$ uptakes under elevated pressures. Such adsorption characteristic is unfavorable in a cyclic process since the adsorbed $CH_4$ result in lower $CH_4$ recovery. Finally, SquCALF-20 shows low $CH_4$ uptakes due to limited pore accessibility of $CH_4$ molecules. We note that these simulation predictions are based on rigid framework assumptions, and the results may vary if we consider the flexibility of the materials.

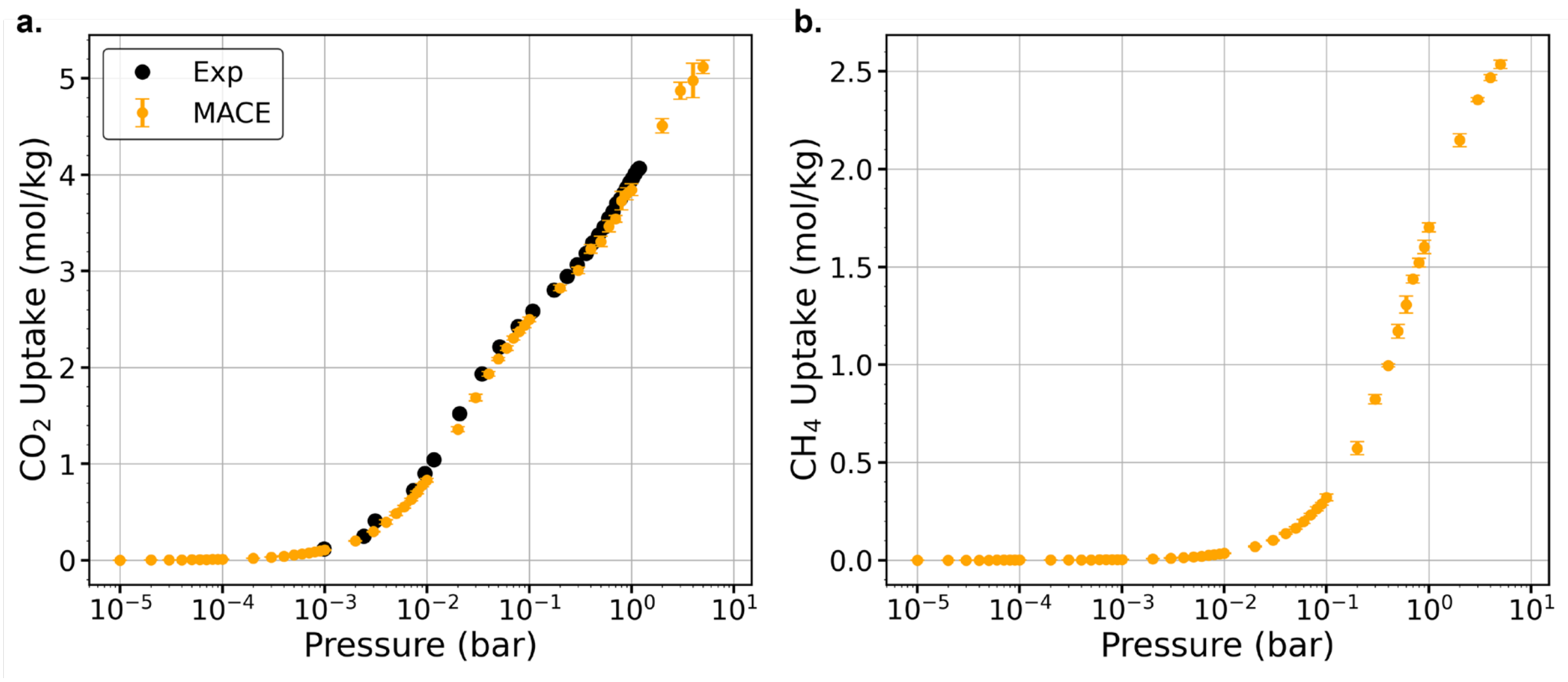

**Fig. 2** Adsorption isotherms of CALF-20 at 298 K obtained from GCMC simulation: **a.** Simulated single-component $CO_2$ isotherm, compared with experimental isotherm data reported by Gopalsamy et al.[36], and **b.** Simulated single-component $CH_4$ isotherm.

Single component GCMC adsorption data of CALF-20 and its derivatives at three temperatures (273, 298, and 323 K) were fitted to the DSL model. The fitted DSL models show excellent agreement with the single-component GCMC simulation results across all materials. Fig. 3 presents the fitting results for CALF-20, while those for the remaining materials – SquCALF-20, FumCALF-20, BdcCALF-20, CubCALF-20, and TtdcCALF-20 – are provided in the Supplementary Information (Figs. S3 and S4). According to the DSL parameters summarized in Table S1, all adsorbent materials exhibited high $CO_2$ adsorption capacities, with TtdcCALF-20 showing the highest $CO_2$ uptake of approximately 12.3 mol/kg. However, TddcCALF-20 demonstrates one of the highest $CH_4$ uptake (~5.5 mol/kg), which may lead to a lower $CH_4$ recovery during adsorption/desorption cycles. In contrast, CALF-20, while showing a comparatively lower $CO_2$ uptake, shows the lowest $CH_4$ adsorption capacity (2.96 mol/kg) among all materials.

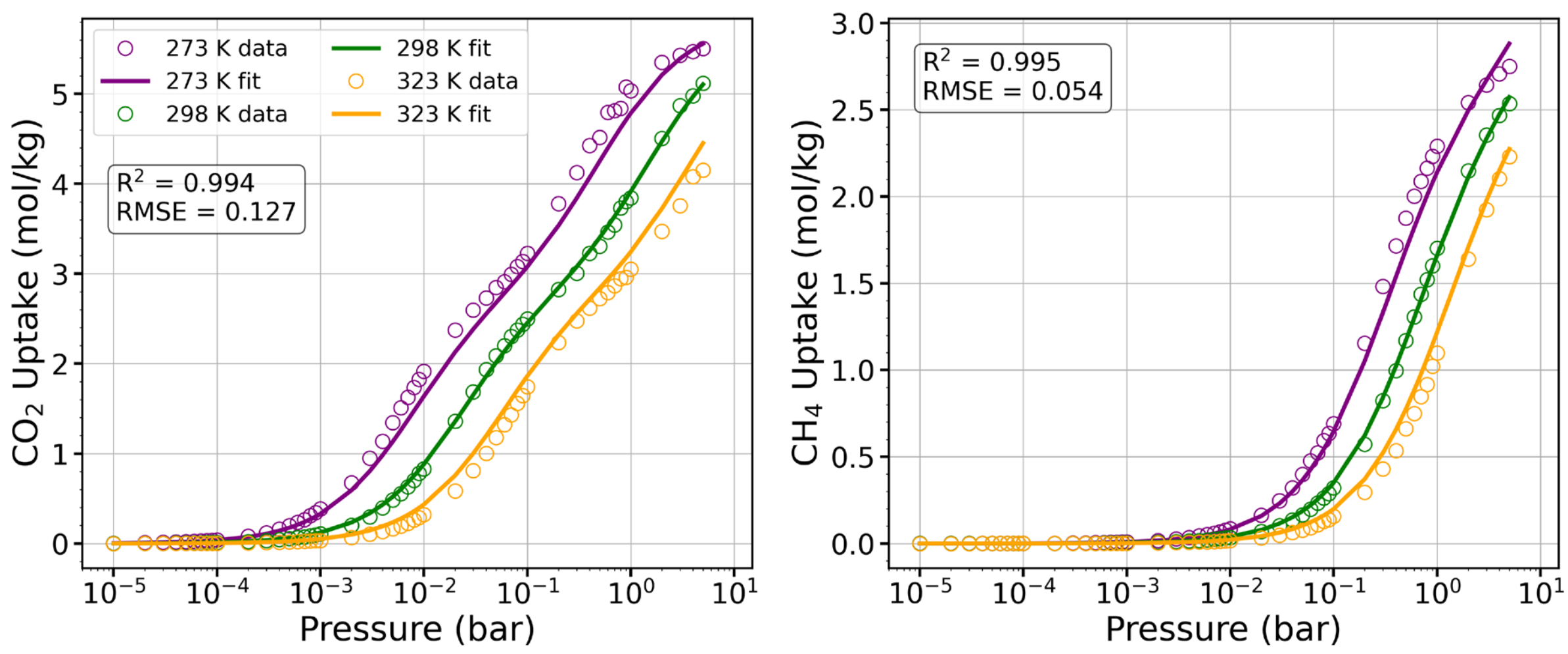

**Fig. 3** Fitting results of **a.** $CO_2$ and **b.** $CH_4$ adsorption isotherms for CALF-20 using the dual-site Langmuir (DSL) model at 273 K, 298 K, and 323 K. Solid lines represent DSL fits, and open symbols denote the corresponding data points.

## 3. Binary mixture adsorption isotherms

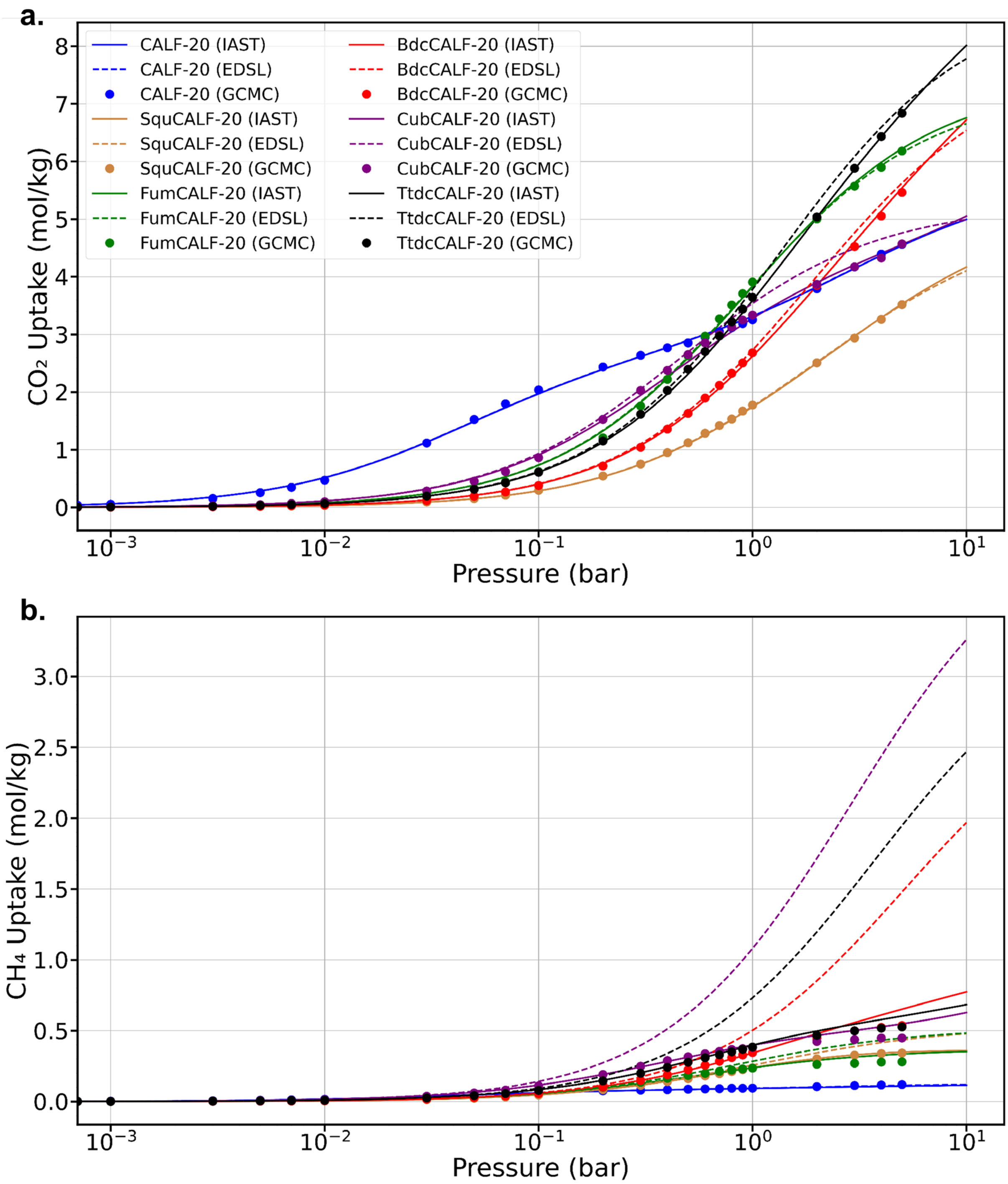

**Fig. 4** Binary mixture ($CH_4$:$CO_2$ = 50:50) adsorption isotherms at 298 K obtained from EDSL and IAST model predictions and GCMC simulations for six CALF-20 derivatives.: **a.** $CO_2$ and **b.** $CH_4$.

The adsorption behavior of $CO_2$/$CH_4$ binary mixtures was analyzed at 298 K with both EDSL and IAST models. As shown in Fig. 4, IAST predictions show excellent agreement with GCMC data for $CO_2$, and the competitive adsorption behavior was reproduced with reasonable accuracy for $CH_4$. While some minor deviations at higher pressures for some derivatives were observed, we found IAST is better at capturing the mixture isotherm behavior than the EDSL fittings (Figs. S5 and S6). All materials showed higher adsorption of $CO_2$ than $CH_4$, indicating a clear preferential adsorption of $CO_2$ even under competitive mixture conditions. At elevated pressures, FumCALF-20 and TtdcCALF-20 exhibit the highest $CO_2$ uptakes, followed by BdcCALF-20. In contrast, SquCALF-20 shows comparatively lower $CO_2$ uptakes. For $CH_4$, CALF-20 presents the lowest uptake, whereas CubCALF-20, TtdcCALF-20, and BdcCALF-20 show relatively high $CH_4$ uptakes, which may adversely affect $CH_4$ recovery during cyclic operation of the process.

The $CO_2$ selectivity and working capacity values under mixture conditions are calculated at the adsorption (1 bar) and desorption (0.1 bar) pressures (Table 5). FumCALF-20 and TtdcCALF-20 show the highest $CO_2$ working capacities, with values of 3.09 mol/kg and 2.96 mol/kg, respectively, followed by CubCALF-20 (2.38 mol/kg) and BdcCALF-20 (2.20 mol/kg). In contrast, SquCALF-20 (1.44 mol/kg) and CALF-20 (1.36 mol/kg) show comparatively lower $CO_2$ working capacities. For $CH_4$, CALF-20 shows the lowest working capacity (0.031 mol/kg), followed by FumCALF-20 (0.184 mol/kg) and SquCALF-20 (0.192 mol/kg). In contrast, BdcCALF-20 (0.283 mol/kg), CubCALF-20 (0.283 mol/kg), and TtdcCALF-20 (0.313 mol/kg) show relatively higher $CH_4$ working capacities, indicating insufficient suppression of $CH_4$ adsorption. In terms of adsorption selectivity, CALF-20 presents the highest $CO_2/CH_4$ ratio (44.0), followed by FumCALF-20 (19.7), TtdcCALF-20 (11.0), CubCALF-20 (10.1), BdcCALF-20 (9.4), and SquCALF-20 (8.9). Table S2 in the Supplementary Information compiles literature data for benchmark sorbents, including zeolites (13X, 5A), activated carbons, and carbon molecular sieves (CMSs). Conventional zeolites such as 13X and 5A typically exhibit $CO_2/CH_4$ selectivities in the range of ~8–20 under similar conditions, whereas activated carbons and CMSs show lower to moderate selectivities of ~3–12 depending on pore structure and preparation. The proposed CALF-20 derivatives achieve comparable or higher $CO_2$ selectivity while maintaining substantial working capacities under PSA-relevant conditions. This comparison highlights that CALF-20 derivatives are competitive with, and in some cases superior to, benchmark sorbents reported in the literature for biogas upgrading.

**Table 5.** Performance metrics based on isotherms: $CO_2$ and $CH_4$ working capacity (WC) and $CO_2/CH_4$ selectivity.

| **Materials** | $\mathbf{WC_{CO_2}}$ **(mol/kg)** | $\mathbf{WC_{CH_4}}$ **(mol/kg)** | $\mathbf{Selectivity_{ads}}$ | $\mathbf{Selectivity_{des}}$ |
|---|---|---|---|---|
| CALF-20 | 1.36 | 0.03 | 44.0 | 39.5 |
| SquCALF-20 | 1.44 | 0.19 | 8.89 | 7.55 |
| FumCALF-20 | 3.09 | 0.18 | 19.7 | 16.7 |
| BdcCALF-20 | 2.20 | 0.28 | 9.35 | 8.63 |
| CubCALF-20 | 2.38 | 0.28 | 10.1 | 9.61 |
| TtdcCALF-20 | 2.96 | 0.31 | 11.0 | 8.92 |

## 4. Ternary mixture analyses

In practical biogas upgrading, it is important to describe adsorption of other gas species, such as nitrogen ($N_2$). The composition of $N_2$ present in biogas stream strongly depends on the feedstock sources. For agricultural and wastewater sources, $N_2$ content is negligible, whereas landfill gas may contain up to ~10% of $N_2$. To investigate the impact of $N_2$ in biogas separation, we conducted ternary GCMC simulations ($CH_4$:$CO_2$:$N_2$ = 50:40:10) at 298 K. The resulting data show negligible $N_2$ adsorption (see Fig. S7) across all materials, validating our approach to treat the biogas as a binary $CO_2$/$CH_4$ mixture for all subsequent analyses.

## 5. Techno-economic analysis

Techno-economic performance of the optimized PVSA cycles was evaluated for CALF-20 and its five isoreticular derivatives. Table 6 summarizes the corresponding operating conditions, process performance indicators, and detailed cost breakdown obtained from the optimization. The feed composition was fixed at $CO_2$/$CH_4$ = 45/55, representing a typical biogas mixture, and a throughput of approximately 1333 $Nm^3/h$ has been reported in the literature for PSA biogas upgrading plants[62]. This data highlights the distinct operating pressures, step durations, and equipment requirements among the materials, as well as their impact on $CH_4$ purity, recovery, and energy consumption.

**Table 6.** Optimized PVSA operating conditions, process indicators, and cost breakdown for CALF-20 and derivatives

| | CALF-20 | SquCALF-20 | FumCALF-20 | BdcCALF-20 | CubCALF-20 | TtdcCALF-20 |
|---|---|---|---|---|---|---|
| *Operating conditions* | | | | | | |
| High pressure, $\boldsymbol{P_H}$ (Pa) | 176,395 | 495 ,891 | 161,700 | 103,693 | 128,475 | 258,309 |
| Low pressure, $\boldsymbol{P_L}$ (Pa) | 49,959 | 41,323 | 36,758 | 18,942 | 37,356 | 30,678 |
| Feed velocity, $\boldsymbol{v_F}$ (m/s) | 0.45 | 0.70 | 0.88 | 0.51 | 0.60 | 0.45 |
| Heavy reflux ratio, $\boldsymbol{\beta_{HR}}$ | 0.36 | 0.48 | 0.09 | 0.06 | 0.16 | 0.32 |
| Light reflux ratio, $\boldsymbol{\alpha_{LR}}$ | 0.82 | 0.63 | 0.48 | 0.38 | 0.88 | 0.51 |
| Adsorption time, $\boldsymbol{t_{ads}}$ (s) | 209.16 | 61.88 | 121.35 | 134.17 | 321.43 | 162.42 |
| Depressurization time, $\boldsymbol{t_{depres}}$ (s) | 60.11 | 47.40 | 26.12 | 21.38 | 47.04 | 45.79 |
| Pressurization time, $\boldsymbol{t_{pres}}$ (s) | 46.89 | 46.27 | 68.93 | 44.21 | 90.64 | 80.35 |
| Column length, $\boldsymbol{L}$ (m) | 3.99 | 3.14 | 5.21 | 5.01 | 4.34 | 4.22 |
| *Process performance indicators* | | | | | | |
| $CH_4$ purity (%) | 98.03 | 97.80 | 97.13 | 97.77 | 98.08 | 97.22 |
| $CH_4$ recovery (%) | 44.8 | 52.65 | 44.65 | 37.49 | 27.32 | 33.63 |
| $CH_4$ energy consumption (kWh/kg $CH_4$) | 9.35 | 39.03 | 13.29 | 34.58 | 15.39 | 32.59 |
| *Train configuration* | | | | | | |
| Number of columns per train, $\boldsymbol{N_{col}}$ | 3 | 3 | 3 | 3 | 3 | 3 |
| Number of depressurization pumps per train, $\boldsymbol{N_{v,Dpres}}$ | 1 | 1 | 1 | 1 | 1 | 1 |
| Number of light reflux pumps per train, $\boldsymbol{N_{v,LR}}$ | 1 | 1 | 1 | 1 | 1 | 1 |
| Number of parallel trains, $\boldsymbol{M}$ | 1 | 1 | 1 | 1 | 1 | 1 |
| *Cost performance* | | | | | | |
| $CH_4$ production cost, $\boldsymbol{C_{CH_4}^{prod}}$ (\$/kg $CH_4$) | **4.313** | **6.121** | **4.802** | **6.863** | **7.029** | **7.249** |
| EAC (\$/kg $CH_4$) | **0.081** | **0.188** | **0.146** | **0.114** | **0.113** | **0.133** |
| Columns | 0.0097 | 0.0054 | 0.0163 | 0.0178 | 0.0186 | 0.02 |
| Compressors for feed | 0.0267 | 0.0846 | 0.0374 | 0.0018 | 0.0204 | 0.0022 |
| Compressors for heavy reflux | 0.0011 | 0.0086 | 0.0008 | 0 | 0.0004 | 0 |
| Vacuum pumps for depressurization | 0.0079 | 0.0089 | 0.0209 | 0.0302 | 0.0211 | 0.0351 |
| Vacuum pumps for light reflux | 0.0029 | 0.0042 | 0.0113 | 0.0178 | 0.007 | 0.0219 |
| Contingency and fee | 0.0087 | 0.0201 | 0.0156 | 0.0122 | 0.0121 | 0.0142 |
| Site developments, auxiliary buildings, off-site, and utilities | 0.0241 | 0.0558 | 0.0434 | 0.0338 | 0.0337 | 0.0396 |
| OPEX ($/kg $CH_4$) | **4.232** | **5.933** | **4.656** | **6.749** | **6.916** | **7.116** |
| Electricity | 0.655 | 2.732 | 0.930 | 2.421 | 1.077 | 2.282 |
| Adsorbent replacement | 0.052 | 0.019 | 0.089 | 0.09 | 0.086 | 0.099 |
| Labor | 1.47 | 1.251 | 1.476 | 1.758 | 2.413 | 1.959 |
| Supervisory | 0.368 | 0.313 | 0.369 | 0.439 | 0.603 | 0.490 |
| Maintenance | 0.061 | 0.141 | 0.109 | 0.085 | 0.085 | 0.100 |
| Operating supplies | 0.012 | 0.028 | 0.022 | 0.017 | 0.017 | 0.020 |
| Administrative overhead | 0.285 | 0.256 | 0.293 | 0.342 | 0.465 | 0.382 |
| Plant overhead | 1.329 | 1.193 | 1.368 | 1.597 | 2.170 | 1.784 |

### 5.1. Cost performance

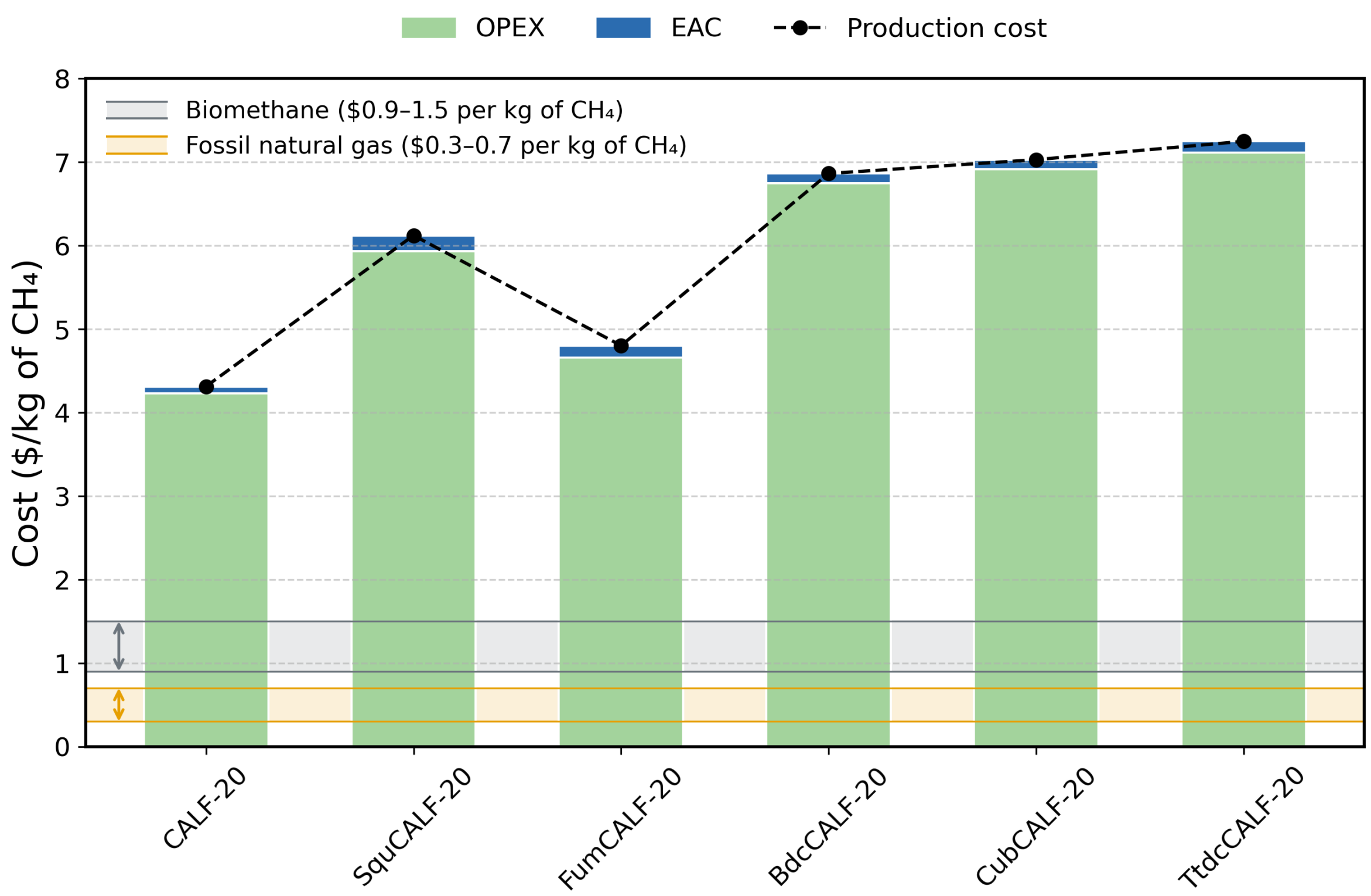

**Fig. 5** Breakdown of $CH_4$ production cost for six CALF-20 derivatives.

Fig. 5 shows the overall $CH_4$ production costs for CALF-20 and other derivatives. Although the $CH_4$ production costs of CALF-20 and FumCALF-20 are competitive relative to other derivatives, they remain considerably higher than the established industrial production cost for biomethane, which typically range from $0.9 to $1.5 per kg of $CH_4$[63]. The total $CH_4$ production cost of this process is predominantly accounted for by operational expenditures (OPEX) while capital expenditures (EAC) account for only about 2–3 % of the total cost.

Fig. 6 shows the breakdown of the OPEX components. The analysis indicates that electricity, labor, and overhead emerged as the major cost contributors. Electricity accounts for a substantial portion of OPEX, reaching 40–50 % in SquCALF-20, BdcCALF-20, and TtdcCALF-20. In contrast, CALF-20 and FumCALF-20 maintained much lower shares (< 20 %), which strongly suggests notable differences in energy efficiency between different materials. Labor was consistently the second-largest contributor, representing 20–35 % of OPEX, reflecting the manpower required for cyclic operation and process control. Overhead costs were also considerable (20–30 % in some cases), which may be associated with higher energy demand or operational complexity. In contrast, the contributions from adsorbent replacement, maintenance, and operating supplies were relatively minor, collectively accounting for less than 5–10 % of OPEX. These results strongly indicate that the overall economics of PVSA processes are influenced more by operational efficiency than by capital investment. The breakdown of the capital expenditures (EAC) is provided in the Supplementary Information (Fig. S8).

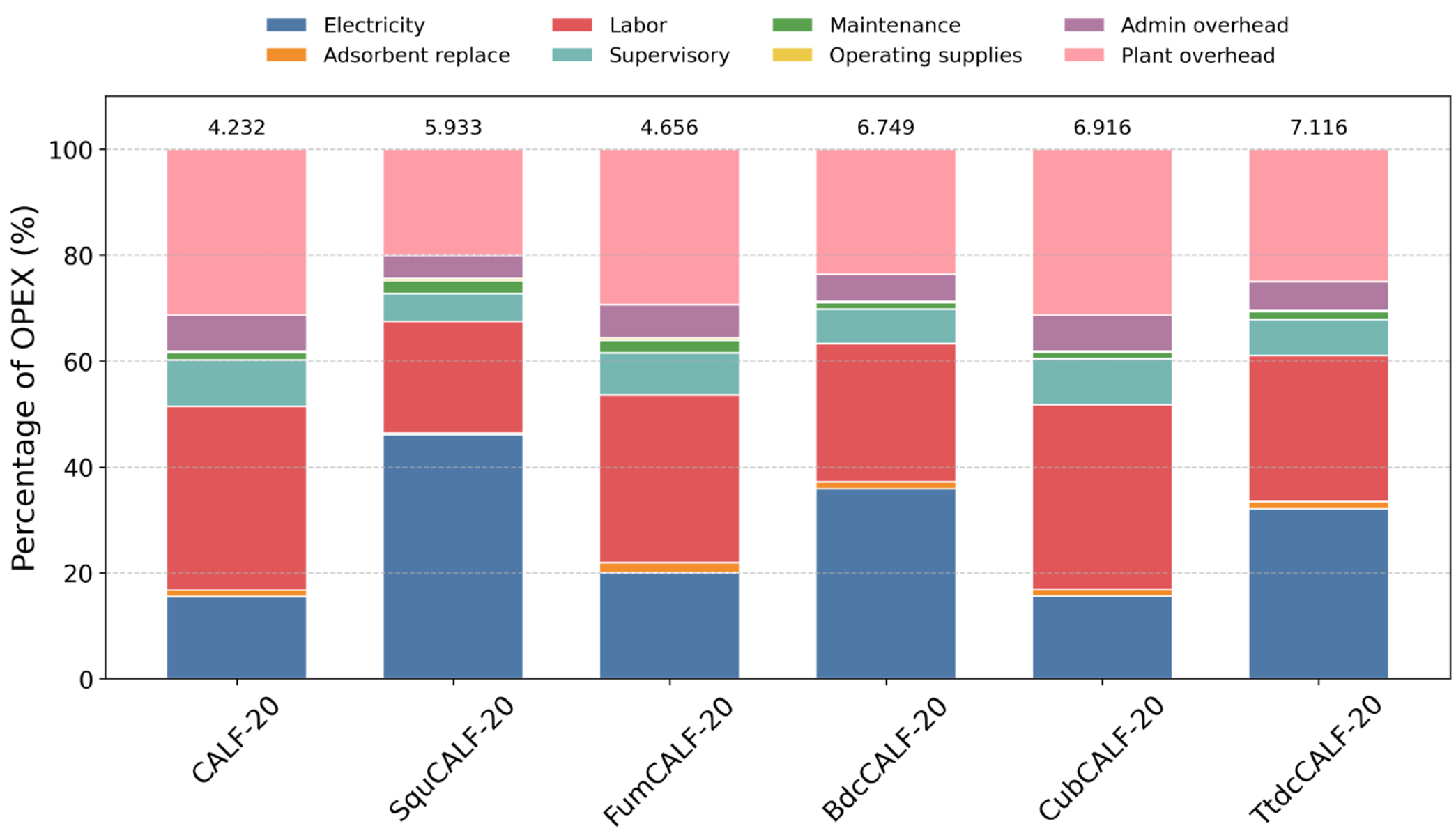

**Fig. 6** Percentage contributions of OPEX to the methane production cost of six CALF-20 derivatives under cost-optimal PVSA operation. The absolute OPEX values (in \$/kg $CH_4$) are indicated above each bar.

Overall, CALF-20 and FumCALF-20 show relatively low total production costs, primarily due to their smaller shares of electricity consumption. This efficiency is attributed to their moderate $CH_4$ working capacities and high $CO_2/CH_4$ selectivities, which are partially reflected in their process-level energy requirements. In contrast, Bdc-, Cub-, and Ttdc-CALF-20 exhibit comparatively large $CH_4$ working capacities and lower $CO_2$ selectivities. This combination may lead to low desorption efficiency and results in less favorable economics. SquCALF-20, despite its low selectivity, achieves higher recovery and thus offers slightly better economic performance than some of the other derivatives, though it remains less competitive than CALF-20 or FumCALF-20. These findings highlight how the adsorption characteristics of the materials are reflected in the process-level energy efficiency and cost calculation, while the structural origins of such techno-economic differences are further discussed in the Supplementary Information (Section S13).

### 5.2. Operating conditions

Table 6 shows the optimized PVSA cycle with distinct operating conditions for different materials. The adsorption pressure ($P_H$) was the highest for SquCALF-20 at 495 kPa, whereas the adsorption pressure for BdcCALF-20 is only 104 kPa. The desorption pressure ($P_L$) was highest for CALF-20 at approximately 50 kPa and lowest for TtdcCALF-20 at 30 kPa, indicating that lower desorption pressures increase vacuum load and electricity demand. These observations demonstrates the differences in $CO_2$ affinity and $CH_4$ desorption efficiency of different materials are directly translated into the optimized adsorption and desorption pressures during cycle simulations. The feed velocity ($v_F$) was greatest for FumCALF-20 at 0.88 m/s, while TtdcCALF-20 exhibited the lowest value of 0.45 m/s. These differences are closely related to the overall throughput of the process and consequently affect both productivity and $CH_4$ production cost. The reflux ratios (heavy/light) also varied substantially: CubCALF-20 shows an extreme distribution (0.16/0.88), suggesting an operating condition more specialized for $CO_2$ removal rather than $CH_4$ recovery. In contrast, FumCALF-20 and CALF-20 show more balanced ratios of 0.48/0.48 and 0.36/0.82, respectively. The cycle step durations (adsorption, depressurization, and repressurization) also differed significantly among the materials. For example,

CubCALF-20 requires the longest adsorption step of 321 s, which reduces throughput and penalizes productivity, whereas SquCALF-20 only requires 62 s, allowing for rapid cycle operation.

### 5.3. Train configuration

The optimization results show that all materials converged to the same train configuration (Fig. S9). Each train required three adsorption columns, and only one vacuum pump was necessary for both the depressurization and light reflux steps. This outcome arises from the structural characteristics of the five-step modified Skarstrom cycle. Since adsorption, heavy reflux, and light reflux must be performed simultaneously, at least three columns are inherently required. In addition, under the optimized step time distributions, the adsorption step was consistently longer than the depressurization step, making it feasible to operate the depressurization stage with a single vacuum pump. The light reflux step could likewise be satisfied with a single pump. Therefore, while all materials show substantial differences in process performance and $CH_4$ production cost, the same train configuration was obtained as an optimal configuration.

### 5.4. Process performance indicators

Under the optimized conditions, all materials achieved $CH_4$ purities above 97%, thereby meeting the pipeline transport requirement (≥ 97%). However, substantial differences were observed in recovery. SquCALF-20 could reach the recovery of 52.7%, whereas CALF-20 and FumCALF-20 could moderate recoveries of 44.8% and 44.7%, respectively. BdcCALF-20 achieved 37.5%, while CubCALF-20 and TtdcCALF-20 recorded the lowest recoveries of 27.3% and 33.6%, indicating potential disadvantages for practical deployment. Electricity consumption also varied sharply across the materials. CALF-20 and FumCALF-20 require only 9.35 and 10.3 kWh per kg $CH_4$, respectively, highlighting their favorable economic performance. By contrast, SquCALF-20 demanded 39.3 kWh/kg $CH_4$, while TtdcCALF-20 and BdcCALF-20 consume 32.6 and 34.6 kWh per kg $CH_4$, more than three times higher than CALF-20. CubCALF-20 shows an intermediate consumption of 15.4 kWh per kg $CH_4$, higher than CALF-20 but lower than the other high-cost materials. In summary, although all materials met the purity constraint, their recoveries and energy demands varied significantly. Among all adsorbents we evaluated,

CALF-20 and FumCALF-20 emerged as economically competitive, combining moderate recovery with excellent energy efficiency. In contrast, SquCALF-20, despite its high recovery, accrued a high $CH_4$ production cost due to its elevated operational energy demands. CubCALF-20 and TtdcCALF-20 exhibited the least favorable combination, suffering from both low recovery and high operational energy demand. The results ultimately demonstrate that beyond meeting the purity requirement, recovery, energy efficiency, and ultimately the production (or separation) cost needs to be considered for ranking the overall economic viability of the adsorbent materials.

## Conclusions

This study conducts a multi-scale computational investigation of CALF-20 and its isoreticular derivatives for biogas upgrading, integrating molecular simulations with PVSA process optimization, followed by techno-economic analyses. CALF-20 emerge as the most economically feasible, with >97% purity $CH_4$ production cost at \$4.31 per kg of $CH_4$ and energy consumption of 9.35 kWh per kg of $CH_4$. This performance stems from the material's low electricity demand and balanced $CH_4$ recovery, which originates from the materials' selective adsorption of $CO_2$ over $CH_4$. Other materials, including SquCALF-20, CubCALF-20, TtdcCALF-20, and BdcCALF-20 suffers from excessive power consumption likely due to decreased selectivity between $CO_2$ and $CH_4$ due to decreased number of preferential adsorption sites for $CO_2$ with ligand modulation. These effects allow $CH_4$ to co-adsorb with $CO_2$, leading to decreased $CH_4$ recovery and high energy costs. The findings indicate that the techno-economic outcomes are governed by the balance between recovery and energy efficiency, a key performance metric fundamentally rooted in the molecular-level adsorption characteristics of adsorbate molecules. We note that although all CALF-20 derivatives achieve $CO_2/CH_4$ separation, their economic performances remain uncompetitive compared to industrial benchmarks. Nevertheless, the evaluation workflow and methodology demonstrated in this work show strong potential to guide the discovery of more economically viable adsorbent materials for biogas upgrading in the future.

## Conflicts of interest

There are no conflicts to declare.

## Acknowledgements

This work was supported by a 2-year research grant from Pusan National University to Yongchul G. Chung.

## References

(1) Balcombe, P.; Speirs, J. F.; Brandon, N. P.; Hawkes, A. D. Methane emissions: choosing the right climate metric and time horizon. *Environmental Science: Processes & Impacts* **2018**, *20* (10), 1323-1339. DOI: 10.1039/C8EM00414E.

(2) Calbry-Muzyka, A.; Madi, H.; Rüsch-Pfund, F.; Gandiglio, M.; Biollaz, S. Biogas composition from agricultural sources and organic fraction of municipal solid waste. *Renewable Energy* **2022**, *181*, 1000-1007. DOI: 10.1016/j.renene.2021.09.100.

(3) Khalil, M.; Berawi, M. A.; Heryanto, R.; Rizalie, A. Waste to energy technology: The potential of sustainable biogas production from animal waste in Indonesia. *Renewable and Sustainable Energy Reviews* **2019**, *105*, 323-331. DOI: 10.1016/j.rser.2019.02.011.

(4) Scarlat, N.; Dallemand, J.-F.; Fahl, F. Biogas: Developments and perspectives in Europe. *Renewable energy* **2018**, *129*, 457-472. DOI: 10.1016/j.renene.2018.03.006.

(5) Leonzio, G. Upgrading of biogas to bio-methane with chemical absorption process: simulation and environmental impact. *Journal of Cleaner Production* **2016**, *131*, 364-375. DOI: 10.1016/j.jclepro.2016.05.020.

(6) Abdeen, F. R.; Mel, M.; Jami, M. S.; Ihsan, S. I.; Ismail, A. F. A review of chemical absorption of carbon dioxide for biogas upgrading. *Chinese Journal of Chemical Engineering* **2016**, *24* (6), 693-702. DOI: 10.1016/j.cjche.2016.05.006.

(7) Baena-Moreno, F. M.; Le Saché, E.; Pastor-Perez, L.; Reina, T. Membrane-based technologies for biogas upgrading: a review. *Environmental Chemistry Letters* **2020**, *18* (5), 1649-1658. DOI: 10.1007/s10311-020-01036-3.

(8) Baena-Moreno, F. M.; Rodríguez-Galán, M.; Vega, F.; Vilches, L. F.; Navarrete, B.; Zhang, Z. Biogas upgrading by cryogenic techniques. *Environmental Chemistry Letters* **2019**, *17*, 1251-1261. DOI: 10.1007/s10311-019-00872-2.

(9) Augelletti, R.; Conti, M.; Annesini, M. C. Pressure swing adsorption for biogas upgrading. A new process configuration for the separation of biomethane and carbon dioxide. *Journal of Cleaner*

*Production* **2017**, *140*, 1390-1398. DOI: 10.1016/j.jclepro.2016.10.013.

(10) Kim, H.; Lee, J.; Lee, S.; Han, J.; Lee, I.-B. Operating optimization and economic evaluation of multicomponent gas separation process using pressure swing adsorption and membrane process. *Korean Chemical Engineering Research* **2015**, *53* (1), 31-38. DOI: 10.9713/kcer.2015.53.1.31.

(11) Abd, A. A.; Othman, M. R.; Helwani, Z.; Kim, J. An overview of biogas upgrading via pressure swing adsorption: Navigating through bibliometric insights towards a conceptual framework and future research pathways. *Energy Conversion and Management* **2024**, *306*, 118268. DOI: 10.1016/j.enconman.2024.118268

(12) Santos, M. P.; Grande, C. A.; Rodrigues, A. E. Dynamic study of the pressure swing adsorption process for biogas upgrading and its responses to feed disturbances. *Industrial & Engineering Chemistry Research* **2013**, *52* (15), 5445-5454. DOI: 10.1021/ie303606v.

(13) Chen, Y.-F.; Lin, P.-W.; Chen, W.-H.; Yen, F.-Y.; Yang, H.-S.; Chou, C.-T. Biogas upgrading by pressure swing adsorption with design of experiments. *Processes* **2021**, *9* (8), 1325. DOI: 10.3390/pr9081325.

(14) Danaci, D.; Bui, M.; Mac Dowell, N.; Petit, C. Exploring the limits of adsorption-based CO 2 capture using MOFs with PVSA–from molecular design to process economics. *Molecular Systems Design & Engineering* **2020**, *5* (1), 212-231. DOI: 10.1039/C9ME00102F.

(15) Santos, M. S.; Grande, C. A.; Rodrigues, A. E. New cycle configuration to enhance performance of kinetic PSA processes. *Chemical Engineering Science* **2011**, *66* (8), 1590-1599. DOI: 10.1016/j.ces.2010.12.032.

(16) Li, Y.; Yi, H.; Tang, X.; Li, F.; Yuan, Q. Adsorption separation of CO2/CH4 gas mixture on the commercial zeolites at atmospheric pressure. *Chemical Engineering Journal* **2013**, *229*, 50-56. DOI: 10.1016/j.cej.2013.05.101.

(17) Karimi, M.; Ferreira, A.; Rodrigues, A. E.; Nouar, F.; Serre, C.; Silva, J. A. MIL-160 (Al) as a candidate for biogas upgrading and CO2 capture by adsorption processes. *Industrial & Engineering Chemistry Research* **2023**, *62* (12), 5216-5229. DOI: 10.1021/acs.iecr.2c04150.

(18) Yan, X.-W.; Bigdeli, F.; Abbasi-Azad, M.; Wang, S.-J.; Morsali, A. Selective separation of

CO2/CH4 gases by metal-organic framework-based composites. *Coordination Chemistry Reviews* **2024**, *520*, 216126. DOI: 10.1016/j.ccr.2024.216126.

(19) Ward, A.; Pini, R. Efficient bayesian optimization of industrial-scale pressure-vacuum swing adsorption processes for co2 capture. *Industrial & Engineering Chemistry Research* **2022**, *61* (36), 13650-13668. DOI: 10.1021/acs.iecr.2c02313.

(20) Hao, Z.; Caspari, A.; Schweidtmann, A. M.; Vaupel, Y.; Lapkin, A. A.; Mhamdi, A. Efficient hybrid multiobjective optimization of pressure swing adsorption. *Chemical Engineering Journal* **2021**, *423*, 130248. DOI: 10.1016/j.cej.2021.130248.

(21) Hartzog, D.; Sircar, S. Sensitivity of PSA process performance to input variables. *Adsorption* **1995**, *1*, 133-151. DOI: 10.1007/BF00705001.

(22) Hao, P.; Shi, Y.; Li, S.; Zhu, X.; Cai, N. Correlations between adsorbent characteristics and the performance of pressure swing adsorption separation process. *Fuel* **2018**, *230*, 9-17. DOI: 10.1016/j.fuel.2018.05.030.

(23) Rosen, A. S.; Iyer, S. M.; Ray, D.; Yao, Z.; Aspuru-Guzik, A.; Gagliardi, L.; Notestein, J. M.; Snurr, R. Q. Machine learning the quantum-chemical properties of metal–organic frameworks for accelerated materials discovery. *Matter* **2021**, *4* (5), 1578-1597. DOI: 10.1016/j.matt.2021.02.015.

(24) Leperi, K. T.; Chung, Y. G.; You, F.; Snurr, R. Q. Development of a general evaluation metric for rapid screening of adsorbent materials for postcombustion CO2 capture. *ACS sustainable chemistry & engineering* **2019**, *7* (13), 11529-11539. DOI: 10.1021/acssuschemeng.9b01418.

(25) Cleeton, C.; de Oliveira, F. L.; Neumann, R. F.; Farmahini, A. H.; Luan, B.; Steiner, M.; Sarkisov, L. A process-level perspective of the impact of molecular force fields on the computational screening of MOFs for carbon capture. *Energy & Environmental Science* **2023**, *16* (9), 3899-3918. DOI: 10.1039/D3EE00858D.

(26) Banu, A.-M.; Friedrich, D.; Brandani, S.; Düren, T. A multiscale study of MOFs as adsorbents in H2 PSA purification. *Industrial & Engineering Chemistry Research* **2013**, *52* (29), 9946-9957. DOI: 10.1021/ie4011035.

(27) Taddei, M.; Petit, C. Engineering metal–organic frameworks for adsorption-based gas

separations: From process to atomic scale. *Molecular Systems Design & Engineering* **2021**, *6* (11), 841-875. DOI: 10.1039/D1ME00085C.

(28) Zhang, N.; Hu, S.; Xin, Q. Optimization of pressure swing adsorption in a three-layered bed for hydrogen purification using machine learning model. *Scientific Reports* **2025**, *15* (1), 14193. DOI: 10.1038/s41598-025-97139-4.

(29) Singh, H. D.; Singh, P.; Vysyaraju, R.; Balasubramaniam, B. M.; Rase, D.; Shekhar, P.; Jose, A.; Rajendran, A.; Vaidhyanathan, R. Unlocking the separation capacities of a 3D-Iron-based metal organic framework built from scarce Fe4O2 Core for upgrading natural gas. *Chemistry of Materials* **2023**, *35* (19), 8261-8271. DOI: 10.1021/acs.chemmater.3c01777.

(30) Karimi, M.; Siqueira, R. M.; Rodrigues, A. E.; Nouar, F.; Silva, J. A.; Serre, C.; Ferreira, A. Biogas upgrading using shaped MOF MIL-160 (Al) by pressure swing adsorption process: Experimental and dynamic modelling assessment. *Separation and Purification Technology* **2024**, *344*, 127260. DOI: 10.1016/j.seppur.2024.127260.

(31) Abd, A. A.; Othman, M. R.; Shamsudin, I. K.; Helwani, Z.; Idris, I. Biogas upgrading to natural gas pipeline quality using pressure swing adsorption for CO2 separation over UiO-66: experimental and dynamic modelling assessment. *Chemical Engineering Journal* **2023**, *453*, 139774. DOI: 10.1016/j.cej.2022.139774.

(32) Rogacka, J.; Pakuła, P.; Mazur, B.; Firlej, L.; Kuchta, B. Advancing biogas purification: A systematic numerical study of MOF performance under humid conditions. *Chemical Engineering Journal* **2024**, *486*, 150097. DOI: 10.1016/j.cej.2024.150097.

(33) Tabar, M. A.; Maghsoudi, H.; Karimi, K.; Hosseini, S. S.; Gholami, M.; Denayer, J. F. Techno-economic analysis of vacuum pressure swing adsorption process for a sustainable upgrading of biogas. *Journal of Cleaner Production* **2024**, *450*, 141853. DOI: 10.1016/j.jclepro.2024.141853.

(34) Ferella, F.; Cucchiella, F.; D'Adamo, I.; Gallucci, K. A techno-economic assessment of biogas upgrading in a developed market. *Journal of Cleaner Production* **2019**, *210*, 945-957. DOI: 10.1016/j.jclepro.2018.11.073.

(35) Oktavian, R.; Goeminne, R.; Glasby, L. T.; Song, P.; Huynh, R.; Qazvini, O. T.; Ghaffari-Nik, O.;

Masoumifard, N.; Cordiner, J. L.; Hovington, P. Gas adsorption and framework flexibility of CALF-20 explored via experiments and simulations. *Nature Communications* **2024**, *15* (1), 3898. DOI: 10.1038/s41467-024-48136-0.

(36) Gopalsamy, K.; Fan, D.; Naskar, S.; Magnin, Y.; Maurin, G. Engineering of an isoreticular series of CALF-20 metal–organic frameworks for CO2 capture. *ACS Applied Engineering Materials* **2024**, *2* (1), 96-103. DOI: 10.1021/acsaenm.3c00622.

(37) Lin, J.-B.; Nguyen, T. T.; Vaidhyanathan, R.; Burner, J.; Taylor, J. M.; Durekova, H.; Akhtar, F.; Mah, R. K.; Ghaffari-Nik, O.; Marx, S. A scalable metal-organic framework as a durable physisorbent for carbon dioxide capture. *Science* **2021**, *374* (6574), 1464-1469. DOI: 10.1126/science.abi7281.

(38) Henle, E. A.; Gantzler, N.; Thallapally, P. K.; Fern, X. Z.; Simon, C. M. PoreMatMod. jl: Julia Package for in Silico Postsynthetic Modification of Crystal Structure Models. *Journal of Chemical Information and Modeling* **2022**, *62* (3), 423-432. DOI: 10.1021/acs.jcim.1c01219.

(39) Module, F. Material Studio 6.0. *Accelrys Inc., San Diego, CA* **2011**.

(40) Rappé, A. K.; Casewit, C. J.; Colwell, K.; Goddard III, W. A.; Skiff, W. M. UFF, a full periodic table force field for molecular mechanics and molecular dynamics simulations. *Journal of the American chemical society* **1992**, *114* (25), 10024-10035.

(41) Larsen, A. H.; Mortensen, J. J.; Blomqvist, J.; Castelli, I. E.; Christensen, R.; Dułak, M.; Friis, J.; Groves, M. N.; Hammer, B.; Hargus, C. The atomic simulation environment—a Python library for working with atoms. *Journal of Physics: Condensed Matter* **2017**, *29* (27), 273002. DOI: 10.1088/1361-648X/aa680e.

(42) Batatia, I.; Benner, P.; Chiang, Y.; Elena, A. M.; Kovács, D. P.; Riebesell, J.; Advincula, X. R.; Asta, M.; Avaylon, M.; Baldwin, W. J. A foundation model for atomistic materials chemistry. *arXiv preprint arXiv:2401.00096* **2023**. DOI: 10.48550/arXiv.2401.00096.

(43) Willems, T. F.; Rycroft, C. H.; Kazi, M.; Meza, J. C.; Haranczyk, M. Algorithms and tools for high-throughput geometry-based analysis of crystalline porous materials. *Microporous and Mesoporous Materials* **2012**, *149* (1), 134-141. DOI: 10.1016/j.micromeso.2011.08.020.

(44) Mayo, S. L.; Olafson, B. D.; Goddard, W. A. DREIDING: a generic force field for molecular

simulations. *Journal of Physical chemistry* **1990**, *94* (26), 8897-8909.
(45) Potoff, J. J.; Siepmann, J. I. Vapor–liquid equilibria of mixtures containing alkanes, carbon dioxide, and nitrogen. *AIChE journal* **2001**, *47* (7), 1676-1682. DOI: 10.1002/aic.690470719.
(46) Martin, M. G.; Siepmann, J. I. Transferable potentials for phase equilibria. 1. United-atom description of n-alkanes. *The Journal of Physical Chemistry B* **1998**, *102* (14), 2569-2577. DOI: 10.1021/jp972543+.
(47) Zhao, G.; Chung, Y. G. PACMAN: A Robust Partial Atomic Charge Predicter for Nanoporous Materials Based on Crystal Graph Convolution Networks. *Journal of chemical theory and computation* **2024**, *20* (12), 5368-5380. DOI: 10.1021/acs.jctc.4c00434.
(48) Widom, B. Some topics in the theory of fluids. *The Journal of Chemical Physics* **1963**, *39* (11), 2808-2812. DOI: 10.1063/1.1734110.
(49) Dubbeldam, D.; Calero, S.; Ellis, D. E.; Snurr, R. Q. RASPA: molecular simulation software for adsorption and diffusion in flexible nanoporous materials. *Molecular Simulation* **2016**, *42* (2), 81-101.
(50) Yancy-Caballero, D.; Leperi, K. T.; Bucior, B. J.; Richardson, R. K.; Islamoglu, T.; Farha, O. K.; You, F.; Snurr, R. Q. Process-level modelling and optimization to evaluate metal–organic frameworks for post-combustion capture of CO 2. *Molecular Systems Design & Engineering* **2020**, *5* (7), 1205-1218. DOI: 10.1039/D0ME00060D.
(51) Haghpanah, R.; Majumder, A.; Nilam, R.; Rajendran, A.; Farooq, S.; Karimi, I. A.; Amanullah, M. Multiobjective optimization of a four-step adsorption process for postcombustion CO2 capture via finite volume simulation. *Industrial & Engineering Chemistry Research* **2013**, *52* (11), 4249-4265. DOI: 10.1021/ie302658y.
(52) Myers, A. L.; Prausnitz, J. M. Thermodynamics of mixed-gas adsorption. *AIChE journal* **1965**, *11* (1), 121-127. DOI: 10.1002/aic.690110125.
(53) Leperi, K. T.; Snurr, R. Q.; You, F. Optimization of two-stage pressure/vacuum swing adsorption with variable dehydration level for postcombustion carbon capture. *Industrial & Engineering Chemistry Research* **2016**, *55* (12), 3338-3350. DOI: 10.1021/acs.iecr.5b03122.
(54) Jiang, G.-S.; Shu, C.-W. Efficient implementation of weighted ENO schemes. *Journal of*

*computational physics* **1996**, *126* (1), 202-228. DOI: 10.1006/jcph.1996.0130.

(55) Zhao, G.; Brabson, L. M.; Chheda, S.; Huang, J.; Kim, H.; Liu, K.; Mochida, K.; Pham, T. D.; Terrones, G. G.; Yoon, S. CoRE MOF DB: A curated experimental metal-organic framework database with machine-learned properties for integrated material-process screening. *Matter* **2025**, *8* (6).

(56) Moosavi, S. M. *tools-cp-porousmat*. GitHub, 2021. https://github.com/SeyedMohamadMoosavi/tools-cp-porousmat (accessed 2025 June 19).

(57) Turton, R.; Shaeiwitz, J. A.; Bhattacharyya, D.; Whiting, W. B. *Analysis, Synthesis, and Design of Chemical Processes*; Prentice Hall, 2018.

(58) Yoon, S.; Jung, J.; Zhao, G.; Lee, Y.; Lee, K. B.; Chung, Y. Techno-Economic Analysis of Cu (I)-Loaded Porous Carbons for Co Separation Using Pressure/Vacuum Swing Adsorption Process. *Vacuum Swing Adsorption Process*. DOI: 10.1016/j.seppur.2025.134786.

(59) Subraveti, S. G.; Roussanaly, S.; Anantharaman, R.; Riboldi, L.; Rajendran, A. Techno-economic assessment of optimised vacuum swing adsorption for post-combustion CO2 capture from steam-methane reformer flue gas. *Separation and Purification Technology* **2021**, *256*, 117832. DOI: 10.1016/j.seppur.2020.117832.

(60) Bradford, E. *TS-EMO*. GitHub, 2020. https://github.com/Eric-Bradford/TS-EMO (accessed 2025 May 8).

(61) Bradford, E.; Schweidtmann, A. M.; Lapkin, A. Efficient multiobjective optimization employing Gaussian processes, spectral sampling and a genetic algorithm. *Journal of global optimization* **2018**, *71* (2), 407-438. DOI: 10.1007/s10898-018-0609-2.

(62) Kohlheb, N.; Wluka, M.; Bezama, A.; Thrän, D.; Aurich, A.; Müller, R. A. Environmental-economic assessment of the pressure swing adsorption biogas upgrading technology. *BioEnergy Research* **2021**, *14* (3), 901-909. DOI: 10.1007/s12155-020-10205-9.

(63) Marconi, P.; Rosa, L. Role of biomethane to offset natural gas. *Renewable and Sustainable Energy Reviews* **2023**, *187*, 113697. DOI: 10.1016/j.rser.2023.113697.

# Supplementary information (SI) for "Multiscale, Techno-economic Evaluation of Isoreticular Series of CALF-20 for Biogas Upgrading using a Pressure/Vacuum Swing Adsorption (PVSA) Process"

Changdon Shin[a], Sunghyun Yoon[a], and Yongchul G. Chung*[ab]

[a]*School of Chemical Engineering, Pusan National University, 46241 Busan, Korea (South)*

[b]*Graduate School of Data Science, Pusan National University, Busan 46241, South Korea*

E-mail: drygchung@gmail.com

This document provides supplementary information for the main manuscript, including detailed adsorption isotherm models, governing equations for PVSA cycle simulations, and adsorption isotherms, and results of isotherm fitting.

- **Section S1. Adsorption Isotherm Models**
- **Section S2. Description of the 5-step modified Skarstrom PVSA cycle**
- **Section S3. Definition of process performance indicators**
- **Section S4. Governing Equations for PVSA Cycle Modeling and Optimization**
- **Section S5. Equations for PVSA train design and parallelization**
- **Section S6. Single component adsorption isotherms from GCMC simulations**
- **Section S7. Curve fitting of $CO_2$ and $CH_4$ isotherms**
- **Section S8. DSL isotherm parameters for $CO_2$ and $CH_4$Section**
- **Section S9. Binary mixture adsorption isotherms**
- **Section S10. Literature data for benchmark $CO_2$/$CH_4$ separation adsorbents**
- **Section S11. Ternary adsorption isotherms from GCMC simulations**
- **Section S12. Additional techno-economic details of the PVSA process**
- **Section S13. Molecular-level adsorption analysis**

**Section S1. Adsorption Isotherm Models**

The adsorption of single components ($CO_2$ and $CH_4$) was modeled using the dual-site Langmuir (DSL) equation, which accounts for two distinct types of adsorption sites with different saturation capacities and binding affinities:

$$q_i = \frac{q^1{}_{sat,i}B^1{}_iP_i}{1+B^1{}_iP_i} + \frac{q^2{}_{sat,i}B^2{}_iP_i}{1+B^2{}_iP_i} \tag{1}$$

where $q_i$ (mmol/g) is the equilibrium adsorption loading of component *i*; $q^s{}_{sat,i}$ (mmol/g) and $B^s{}_i$ (1/Pa) denote the saturation loading, which is the maximum amount of component *i* that can be adsorbed at site *s*, and the affinity between the adsorbent and the adsorbate, respectively; $P_i$ (Pa) represents the pressure of component *i*. Since the affinity constant $B^s{}_i$ is temperature-dependent, we could calculate the value using the van't Hoff equation, as shown below:

$$B^s{}_i = b^s{}_i exp\left(\frac{-\Delta U^s{}_i}{RT}\right) \tag{2}$$

where $\Delta U^s{}_i$ is the change in internal energy of component *i* at site *s*.

For binary mixture adsorption, the Extended dual-site Langmuir (EDSL) model is used:

$$q_i = \frac{q^1{}_{sat,i}B^1{}_iP_i}{1+\sum_j B^1{}_jP_j} + \frac{q^2{}_{sat,i}B^2{}_iP_i}{1+\sum_j B^2{}_jP_j} \tag{3}$$

In this equation, $q_i$, $q^s{}_{sat,i}$, $B^s{}_i$, and $P_i$ retain the same meanings as in the DSL model, extended here to account for the competitive adsorption among multiple components *j* ($CH_4/CO_2$) on each adsorption site *s*.

The mixture adsorption equilibria were also evaluated using the Ideal Adsorbed Solution Theory (IAST). Within this framework, the reduced grand potential of component *i* is given as

$$\psi_i^* = \int_0^{P_i^*} \frac{q_i^0(P)}{P} dP \tag{4}$$

where $P_i^*$ is the fictitious pressure and $q_i^0(P)$ corresponds to the unary adsorption isotherm of component *i*. Thermodynamic equilibrium is achieved when the reduced grand potentials of all components are equal,

$$\psi_1^* = \psi_2^* = \cdots = \psi_N^* \tag{5}$$

and the fictitious pressure is related to the bulk partial pressure and the adsorbed-phase mole fraction according to

$$P_i = x_i P_i^*, \ \ \sum_{i=1}^{N} x_i = 1 \tag{6}$$

## Section S2. Description of the 5-step modified Skarstrom PVSA cycle

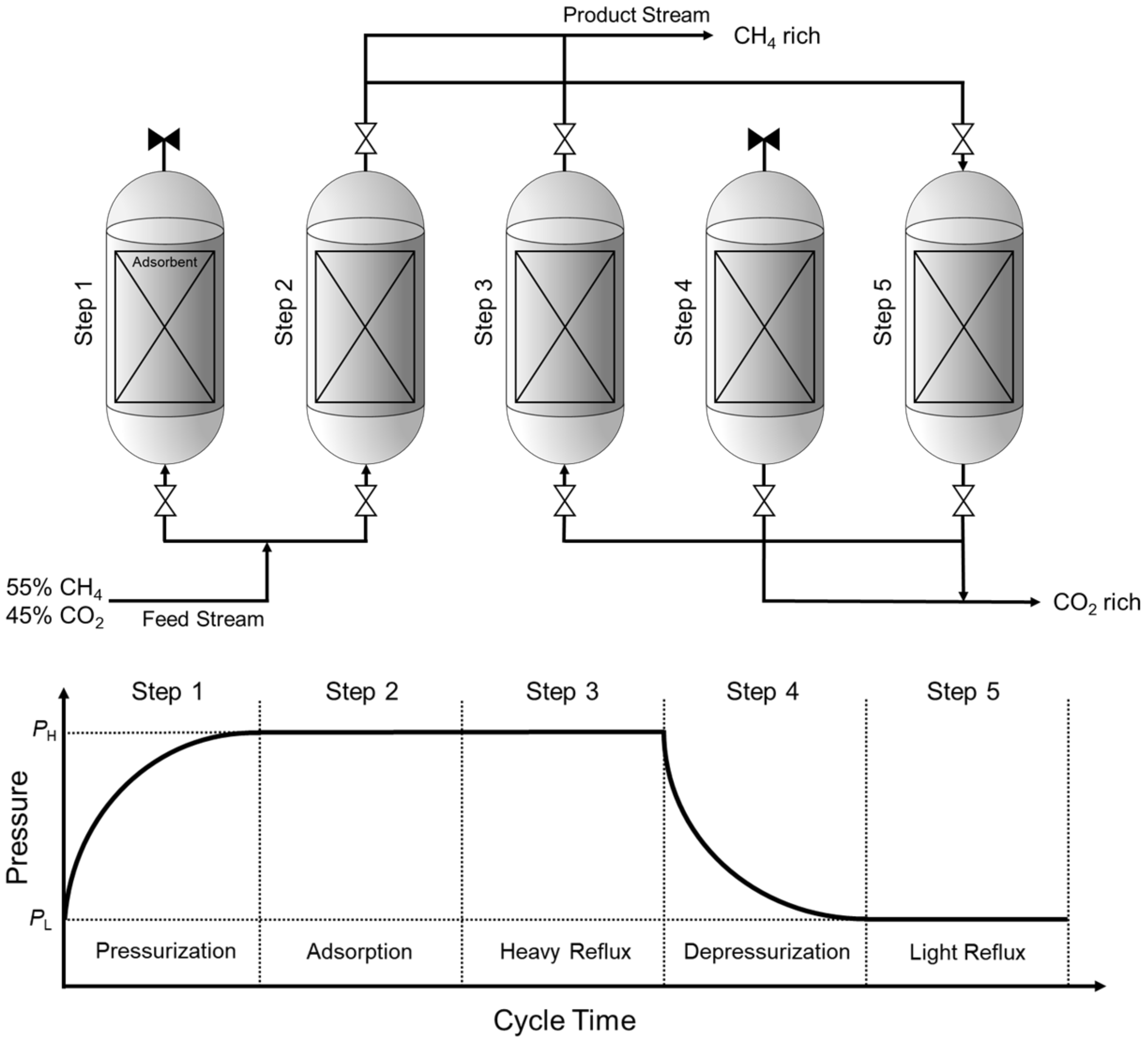

**Scheme S1** Schematic representation of the 5-step modified Skarstrom PVSA cycle. $P_L$ and $P_H$ are the desorption and adsorption pressures, respectively.

The modified Skarstrom cycle employed in this study consists of five sequential steps:

**1. Pressurization**: The feed gas at low pressure ($P_L$) is introduced at the bottom of the column, increasing the pressure to the adsorption pressure ($P_H$).

**2. Adsorption**: The feed gas continues to flow in while the top of the column remains open. $CO_2$ is adsorbed inside the column, and $CH_4$-rich gas exits from the top of the column.

**3. Heavy reflux**: The incoming gas is switched from feed gas to the heavy product generated in the

light reflux step. Additional $CH_4$-rich gas exits from the top of the column.

**4. Counter-current depressurization**: The top of the column is closed, and the pressure is reduced to the desorption pressure ($P_L$). During this step, $CO_2$ desorbs and exits through the bottom inlet of the column.

**5. Light reflux**: After complete depressurization, the light product ($CH_4$-rich gas) obtained in the adsorption step is introduced into the top of the column. Additional $CO_2$-rich gas exits from the bottom of the column, with a portion being used as the feed gas in the heavy reflux step.

### Section S3. Definition of process performance indicators

$CH_4$ purity and recovery were defined as follows:

$$Purity_{CH_4} = \frac{n_{CH_4,out}^{ads} \times (1-\alpha_{LR}) + n_{CH_4,out}^{HR}}{n_{total,out}^{ads} \times (1-\alpha_{LR}) + n_{total,out}^{HR}} \tag{7}$$

$$Recovery_{CH_4} = \frac{n_{CH_4,out}^{ads} \times (1-\alpha_{LR}) + n_{CH_4,out}^{HR}}{n_{CH_4,in}^{pres} + n_{CH_4,in}^{ads}} \tag{8}$$

$n_{CH_4,out}^{ads}$ and $n_{CH_4,out}^{HR}$ are the moles of $CH_4$ exiting during the adsorption and heavy reflux stages, respectively; $n_{total,out}^{ads}$ and $n_{total,out}^{HR}$ are the total moles of gas exiting during the adsorption and heavy reflux stages, respectively; Additionally, $n_{CH_4,in}^{pres}$ and $n_{CH_4,in}^{ads}$ are the moles of $CH_4$ fed into the system during the pressurization and adsorption stages, respectively. The method for calculating energy consumption is as follows:

$$Energy\ Consumption = \frac{E_{total}}{m_{CO_2,out}^{Ads} \times (1-\alpha_{LR}) + m_{CO_2,out}^{HR}} \tag{9}$$

$n_{CO_2,out}^{ads}$ and $n_{CO_2,out}^{HR}$ are the moles of $CO_2$ exiting during the adsorption and heavy reflux stages, respectively; $m_{adsorbent}$ is the mass of the adsorbent, and $t_{cycle}$ is the cycle operating time; $E_{total}$ is the total energy requirement of the cycle, while $m_{CO_2,out}^{ads}$ and $m_{CO_2,out}^{HR}$ are the mass of $CO_2$ exiting during the adsorption and heavy reflux stages, respectively.

**Section S4. Governing Equations for PVSA Cycle Modeling and Optimization**

This section summarizes the set of governing equations used in the dynamic simulation of the PVSA cycle. The model accounts for multicomponent mass transfer, energy balances, and pressure drop within the column under non-isothermal and non-isobaric conditions. Each equation is expressed in its dimensional or nondimensional form as used in the finite-volume simulation framework described in the main text.

Overall mass balance: $\frac{\partial \bar{P}}{\partial \tau} - \frac{\bar{P}}{\bar{T}}\frac{\partial \bar{T}}{\partial \tau} = -\bar{T} - \psi \bar{T} \sum_{i=1}^{n_{comp}} \frac{\partial \bar{x}_i}{\partial \tau}$ (10)

Component mass balance: $\frac{\partial y_i}{\partial \tau} + \frac{y_i}{\bar{P}}\frac{\partial \bar{P}}{\partial \tau} - \frac{y_i}{\bar{T}}\frac{\partial \bar{T}}{\partial \tau} = \frac{1}{\mathrm{Pe}}\frac{\bar{T}}{\bar{P}}\frac{\partial}{\partial Z}\left(\frac{\bar{P}}{\bar{T}}\frac{\partial y_i}{\partial Z}\right) - \frac{\bar{T}}{\bar{P}}\frac{\partial}{\partial Z}\left(\frac{y_i \bar{P} \bar{v}}{\bar{T}}\right) - \frac{\bar{T}}{\bar{P}}\psi \frac{\partial x_i}{\partial \tau}$ (11)

Solid-phase balance: $\frac{\partial x_i}{\partial \tau} = \alpha_i ({x_i}^* - x_i)$ (12)

Pressure drop: $-\frac{\partial \bar{P}}{\partial Z} = \frac{150}{4{r_p}^2}\left(\frac{1-\epsilon}{\epsilon}\right)^2 \frac{v_0 L}{P_0} \mu \bar{v}$ (13)

Column energy balance: $\frac{\partial \bar{T}}{\partial \tau} + \Omega_2 \frac{\partial \bar{P}}{\partial \tau} = \Omega_2 \frac{\partial^2 \bar{T}}{\partial Z^2} - \Omega_2 \frac{\partial}{\partial Z}(\bar{P}\bar{v}) + \sum_{i=1}^{n_{comp}} \left[(\sigma_i - \Omega_3 \bar{T})\frac{\partial x_i}{\partial \tau}\right] - \Omega_4 (\bar{T} - \bar{T}_w)$ (14)

Wall energy balance: $\frac{\partial \bar{T}_w}{\partial \tau} = \Pi_1 \frac{\partial^2 \bar{T}_w}{\partial Z^2} + \Pi_1 (\bar{T} - \bar{T}_w) - \Pi_3 (\bar{T}_w - \bar{T}_a)$ (15)

**Section S5. Equations for PVSA train design and parallelization**

**S5.1. Unit train design**

The minimum number of vacuum pumps required for counter-current depressurization and light reflux in each train:

$$N_{v,j} = ceiling\left(\frac{t_j}{t_{pres} + t_{ads}}\right) \; (j = depres \; or \; LR)$$

where $N_{v,j}$ represents the minimum number of vacuum pumps required for the counter-current depressurization or the light reflux step in each train. After determining the auxiliary equipment requirements, the number of columns in a unit train is obtained. The minimum number of columns per train was calculated as follows:r

$$N_{col} = ceiling\left(\frac{\sum_{i=steps} t_i}{t_{pres} + t_{ads}}\right)$$

where $N_{col}$ represents the minimum number of columns required in a train. If the total duration of the cycle steps is not an exact multiple of the combined pressurization and adsorption times, an additional idle step is introduced. This idle step is inserted after the depressurization stage to minimize disturbances to the bed profile, and its duration is calculated as:

$$t_{idle} = N_{col}\left(t_{pres} + t_{ads}\right) - \sum_{i=steps} t_i$$

where $t_{idle}$ is the duration (s) of the idle step.

**S5.2. Parallel unit trains**

The average molar flow of methane entering one train, $\dot{n}_{CH_4,in}$ is first evaluated over the pressurization and adsorption intervals, is calculated as follows:

$$\dot{n}_{CH_4,in} = \frac{1}{t_{pres} + t_{ads}}\left(\int_0^{t_{pres}} \dot{n}_{CH_4,pres} dt + \int_0^{t_{ads}} \dot{n}_{CH_4,ads} dt\right)$$

where $\dot{n}_{CH_4,pres}$ is the inlet molar flow rate (mol $CH_4$/s) during the pressurization step, and $\dot{n}_{CH_4,ads}$ is the inlet molar flow rate (mol $CH_4$/s) during the adsorption step. The total number of parallel trains, $M$, required to process the full $CO_2$/$CH_4$ feed is then calculated as follows:

$$M = ceiling\left(\frac{\dot{n}_{CH_4,total}}{\dot{n}_{CH_4,in}}\right)$$

where $\dot{n}_{CH_4,total}$ is the total molar flow rate (mol $CH_4$/s) of the $CO_2$/$CH_4$ feed. Accordingly, the full-scale PVSA system comprises $M$ identical trains, each containing $N_{col}$ columns, $N_{v,depres}$ vacuum pumps for depressurization, $N_{v,LR}$ vacuum pumps for light reflux, one feed compressor, and one heavy-reflux compressor.

**S5.3. Cost model**

**S5.3.1. Capital cost**

The purchase cost of an adsorption column can be estimated using the following correlation:

$$\log_{10} C_{P,col} = 3.4974 + 0.4485\log_{10}V + 0.1074(\log_{10}V)^2$$

where $C_{P,col}$ denotes the purchase cost ($) of the adsorption column, and $V$ is the column volume ($m^3$), calculated as follows:

$$V = \pi r_{in}^2 L$$

where $L$ is the column length (m). The BMC of the adsorption column is then calculated as follows:

$$C_{BM,col} = (2.25 + 1.82F_m F_P)C_{P,col} \times \frac{CEPCI_{2024}}{CEPCI_{2001}}$$

where $C_{BM,col}$ is the BMC of the adsorption column, $CEPCI_i$ is the Chemical Engineering Plant Cost Index for year $i$, $F_m$ is the material factor, and $F_P$ is the pressure factor. For an adsorption column constructed of carbon steel, the material factor ($F_m$) is taken as 1, while the pressure factor ($F_P$) is calculated as:

$$F_P = \max\left(\frac{\frac{(P_H+1)2r_{in}}{2\left(850-0.6(P_H+1)\right)}+0.00315}{0.0063}, 1.25\right)$$

where $P_H$ is the adsorption pressure (barg). The purchase cost of the vacuum pump is given by the following correlation:

$$\log_{10} C_{P,v} = 3.3892 + 0.0536\log_{10}\dot{W} + 0.1536\left(\log_{10}\dot{W}\right)^2$$

where $C_{P,v}$ is the purchase cost ($) of the vacuum pump, and $\dot{W}$ is its maximum shaft power (kW). The BMC of the vacuum pump is calculated as follows:

$$C_{BM,v} = (1.89 + 1.35F_m F_P)C_{P,v} \times \frac{CEPCI_{2024}}{CEPCI_{2001}}$$

where $C_{BM,v}$ is the bare module cost ($) of the vacuum pump. For a vacuum pump constructed of carbon steel, the material factor ($F_m$) and pressure factor ($F_P$) are taken as 1.6 and 1, respectively. The total cost of the compressor is estimated as the sum of the compressor body and its electric motor, with their purchase costs expressed by the following correlations:

$$\log_{10} C_{P,comp} = 2.2897 + 1.3604\log_{10}\eta\dot{W} + 0.1027\left(\log_{10}\eta\dot{W}\right)^2$$

$$\log_{10} C_{P,drive} = 1.956 + 1.7142\log_{10}\dot{W} + 0.2282\left(\log_{10}\dot{W}\right)^2$$

where $C_{P,comp}$ is the purchase cost ($) of the compressor, $C_{P,drive}$ is the purchase cost ($) of the electric motor, and $\eta$ is the efficiency of the compressor, and $\dot{W}$ is the maximum shaft power (kW) of the compressor. The BMC of the compressor and its drive are then calculated as follows:

$$C_{BM,comp} = 2.7C_{P,comp} \times \frac{CEPCI_{2024}}{CEPCI_{2001}}$$

$$C_{BM,drive} = 1.5C_{P,drive} \times \frac{CEPCI_{2024}}{CEPCI_{2001}}$$

where $C_{BM,comp}$ is the BMC ($) of the compressor, and $C_{BM,drive}$ is the BMC ($) of the electric motor.

The total bare module cost (TBMC) is subsequently obtained as the sum of the BMCs of all equipment, calculated as follows:

$$TBMC(\$) = MN_{col}C_{BM,col} + MN_{v,depres}C_{BM,v,depres} + MN_{v,LR}C_{BM,v,LR} + M(C_{BM,comp,feed} + C_{BM,drive,feed}) + M(C_{BM,comp,HR} + C_{BM,drive,HR})$$

where $C_{BM,v,k}$ is the BMC (\$) of the vacuum pump used in the counter-current depressurization or light reflux step, and $C_{BM,comp,k}$ is the BMC (\$) of the compressor used in the feed or heavy reflux step. Contingency and fee costs were assumed to be 15% and 3% of the TBMC, respectively. By incorporating these costs into the TBMC, the total module cost (TMC) is calculated as follows:

$$TMC(\$) = 1.18 \times TBMC$$

A total of 50% of the TBMC was assumed to account for additional costs related to site development, auxiliary buildings, off-site facilities, and utilities. By adding these costs to the TMC, the CAPEX is calculated as follows:

$$CAPEX(\$) = TMC + 0.50 \times TBMC$$

The CAPEX was annualized using the equivalent annual cost (EAC) approach[1], and the annualized CAPEX, or EAC, is calculated as follows:

$$EAC(\$/yr) = \frac{CAPEX \times d}{1 - (1 + d)^{-t}}$$

Where $d$ is the discount rate and $t$ is the economic lifetime of the project.

**S5.3.2. Operating cost**

The electricity cost associated with the operation of vacuum pumps and compressors is calculated as follows:

$$OC_{elec} = E_{total}RE_{CH_4}\dot{m}_{CH_4,total}C_{elec}$$

where $OC_{elec}$ is the annual electricity cost for operating the vacuum pumps and compressors (\$/yr),

$\dot{m}_{CH_4,total}$ is the total mass flow rate (tonne $CH_4$/yr) of the $CO_2/CH_4$ feed, and $C_{elec}$ is the unit electricity cost ($/kWh). Assuming that the adsorbent is completely replaced every 1.5 years, the corresponding replacement cost is calculated as follows

$$OC_{ads} = MN_{elec}V(1-\varepsilon)\rho_s C_{ads}/1.5$$

where $OC_{ads}$ is the annual cost of adsorbent replacement ($/yr), and ε is the column void fraction, and $\rho_s$ is the adsorbent density (kg/m³). The cost of the adsorbent ($C_{ads}$) was assumed to be $10 per kg[2]. For labor cost estimation, it was assumed that 10 operators, including technicians, are required to operate the PVSA system at a labor rate of $34.50 per hour. The labor cost is calculated as follows:

$$OC_{labor} = Laborrate \times 10 \times 365 \times 24$$

where $OC_{labor}$ is the labor cost ($/yr). The total operating cost, or OPEX, is calculated as the sum of these components, as follows:

$$OPEX(\$/yr) = OC_{elec} + OC_{ads} + OC_{labor} + OC_{supervisory} + OC_{maintenance} + OC_{supplies} + OC_{admin} + OC_{plant}$$

The supervisory cost ($OC_{supervisory}$) is calculated as 25% of the labor cost, the maintenance cost ($OC_{maintenance}$) as 10% of the TMC, and the operating supplies ($OC_{supplies}$) as 20% of the maintenance cost. The administrative overhead ($OC_{admin}$) and plant overhead ($OC_{plant}$) are calculated as 15% and 70%, respectively, of the sum of labor, supervisory, and maintenance costs.

### S5.3.3. $CH_4$ production cost

The total annual cost (TAC) is calculated as the sum of the annualized CAPEX (EAC) and the OPEX, as follows:

$$TAC(\$/yr) = EAC \times OPEX$$

The $CH_4$ production cost is determined based on the TAC, as follows:

$$C_{CH_4}^{prod}(\$/tonne\ CH_4) = \frac{TAC}{RE_{CH_4}\dot{m}_{CH_4,total}}$$

**Section S6. Single component adsorption isotherms from GCMC simulations**

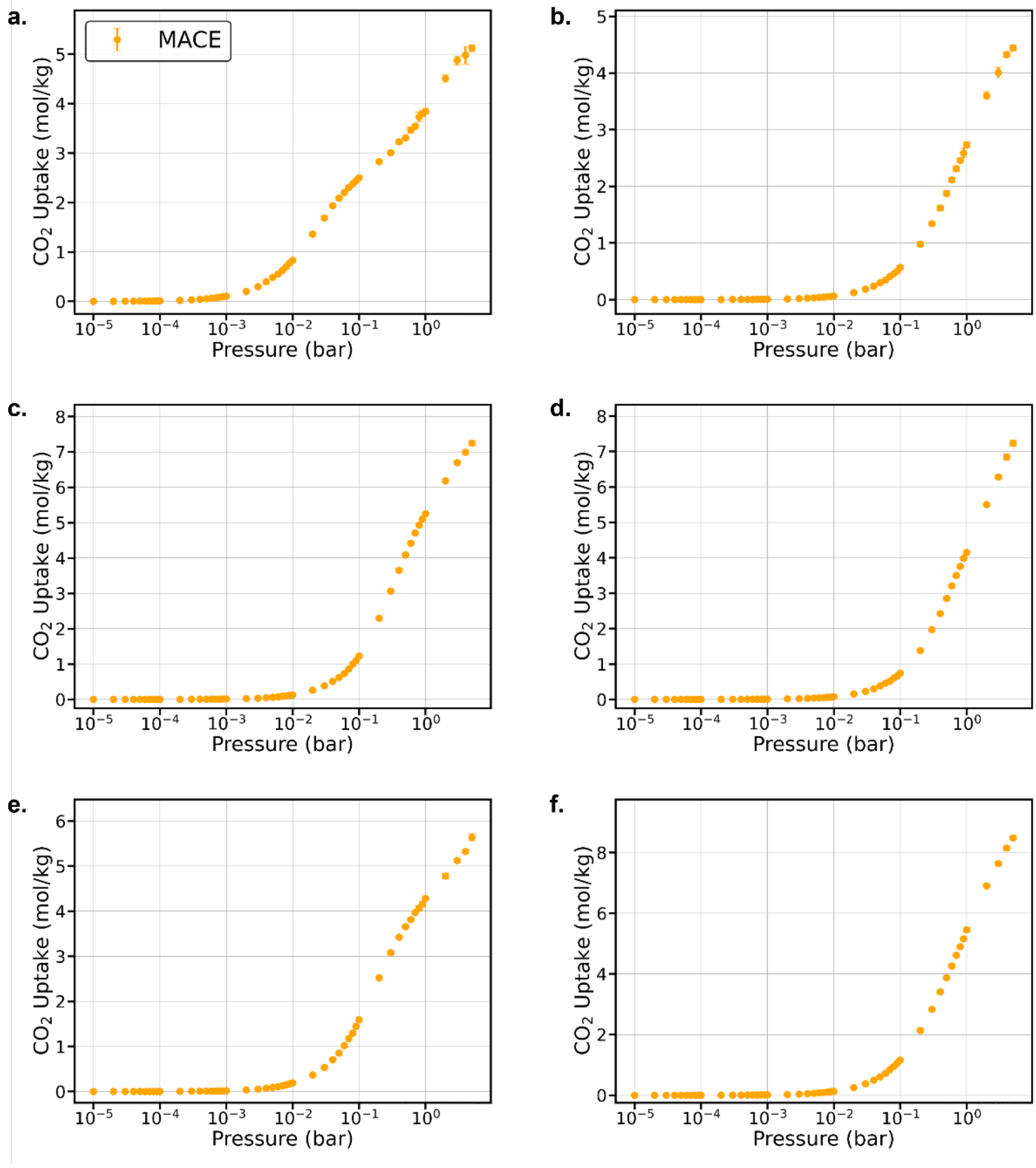

**Fig. S1** $CO_2$ adsorption isotherms in **a.** CALF-20, **b.** SquCALF-20, **c.** FumCALF-20, **d.** Bdc-CALF-20, **e.** CubCALF-20, and **f.** Ttdc-CALF-20 obtained from GCMC simulations at 298 K.

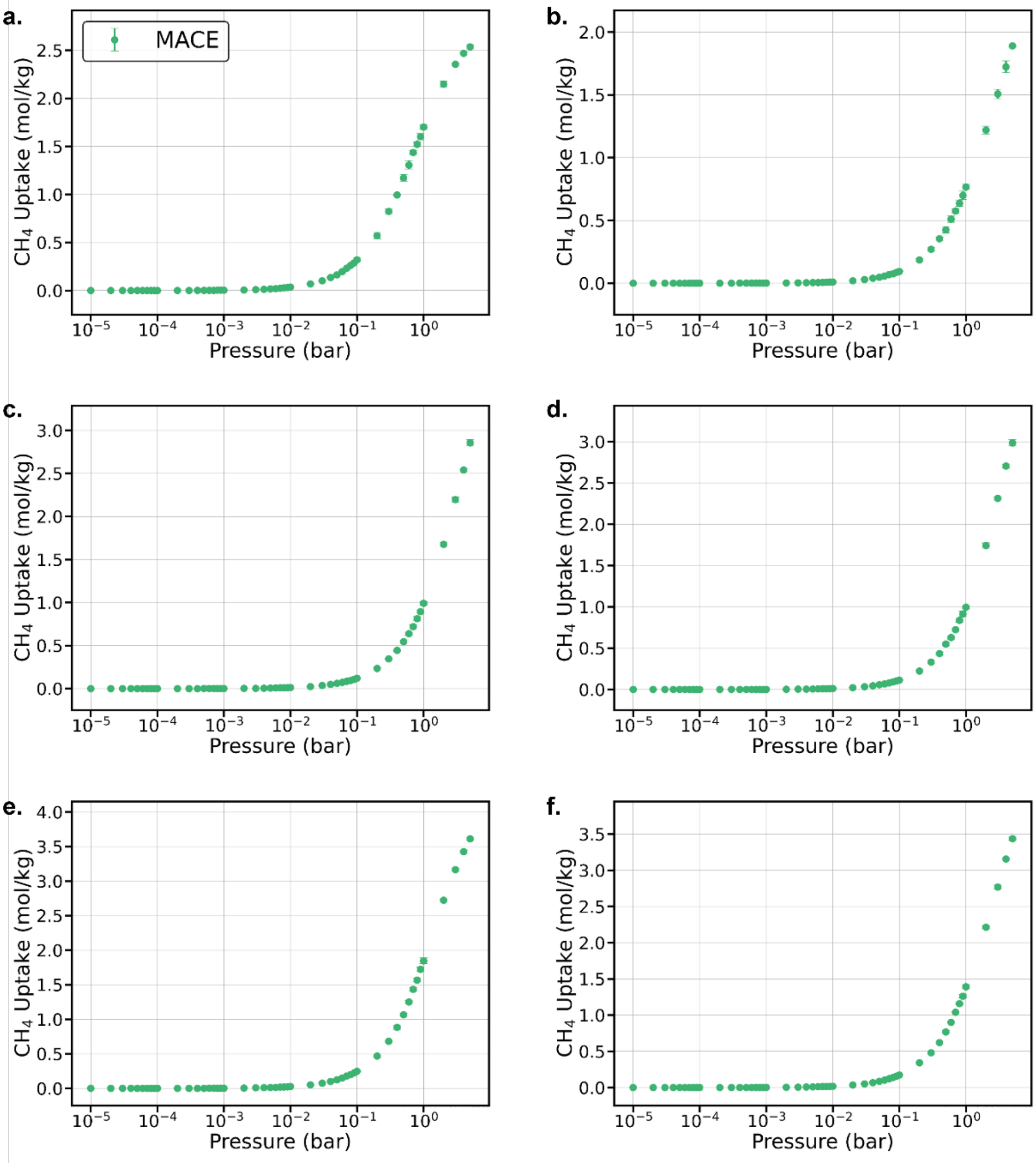

**Fig. S2** $CH_4$ adsorption isotherms in **a.** CALF-20, **b.** SquCALF-20, **c.** FumCALF-20, **d.** Bdc-CALF-20, **e.** CubCALF-20, and **f.** Ttdc-CALF-20 obtained from GCMC simulations at 298 K.

**Section S7. Curve fitting of $CO_2$ and $CH_4$ isotherms**

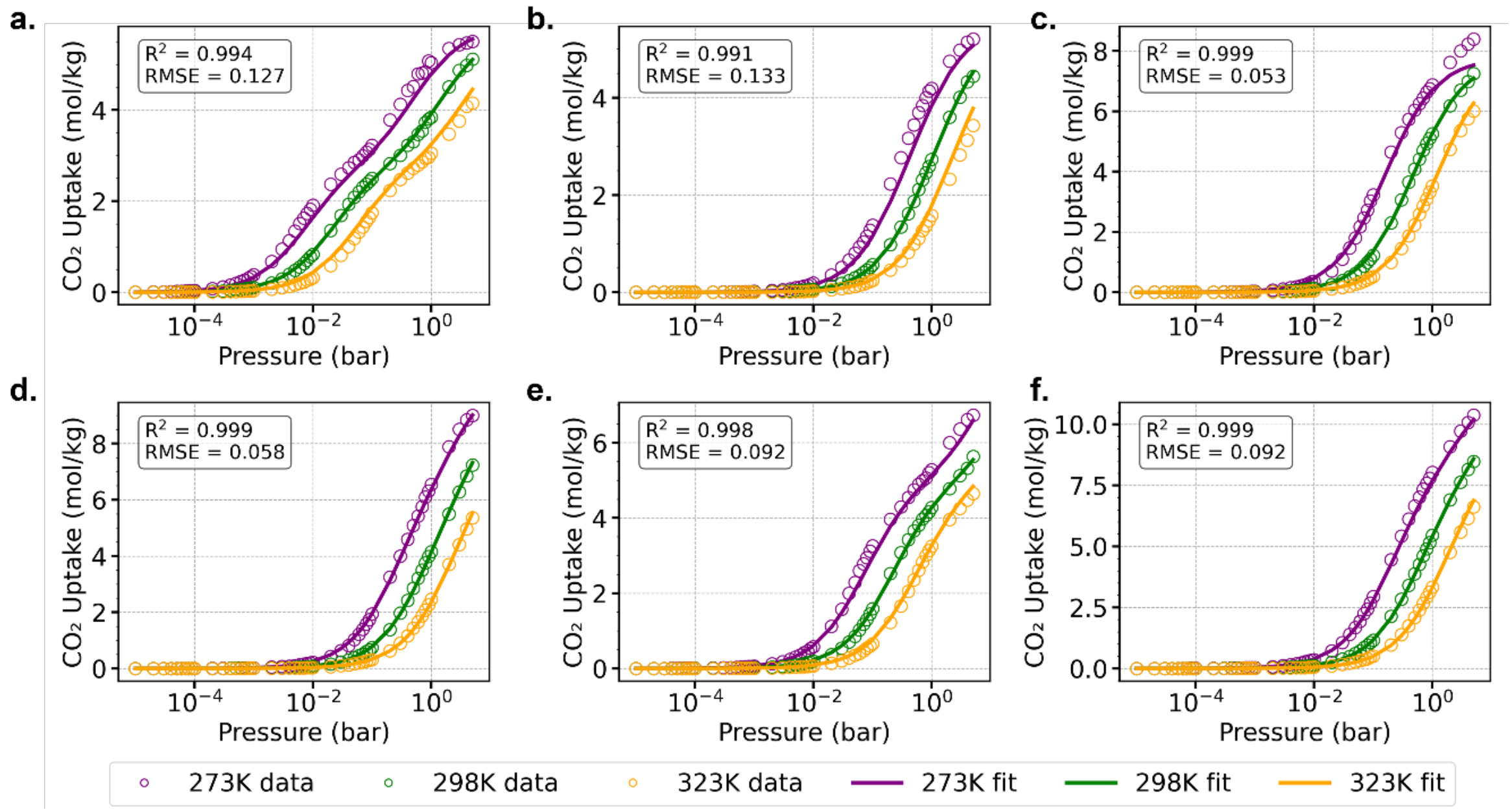

**Fig. S3** Curve fitting results of $CO_2$ isotherms for **a.** CALF-20, **b.** SquCALF-20, **c.** FumCALF-20, **d.** Bdc-CALF-20, **e.** CubCALF-20, and **f.** Ttdc-CALF-20 using the dual-site Langmuir (DSL) model at 273 K, 298 K, and 323 K. Solid lines represent DSL fits, and open symbols denote the corresponding data points.

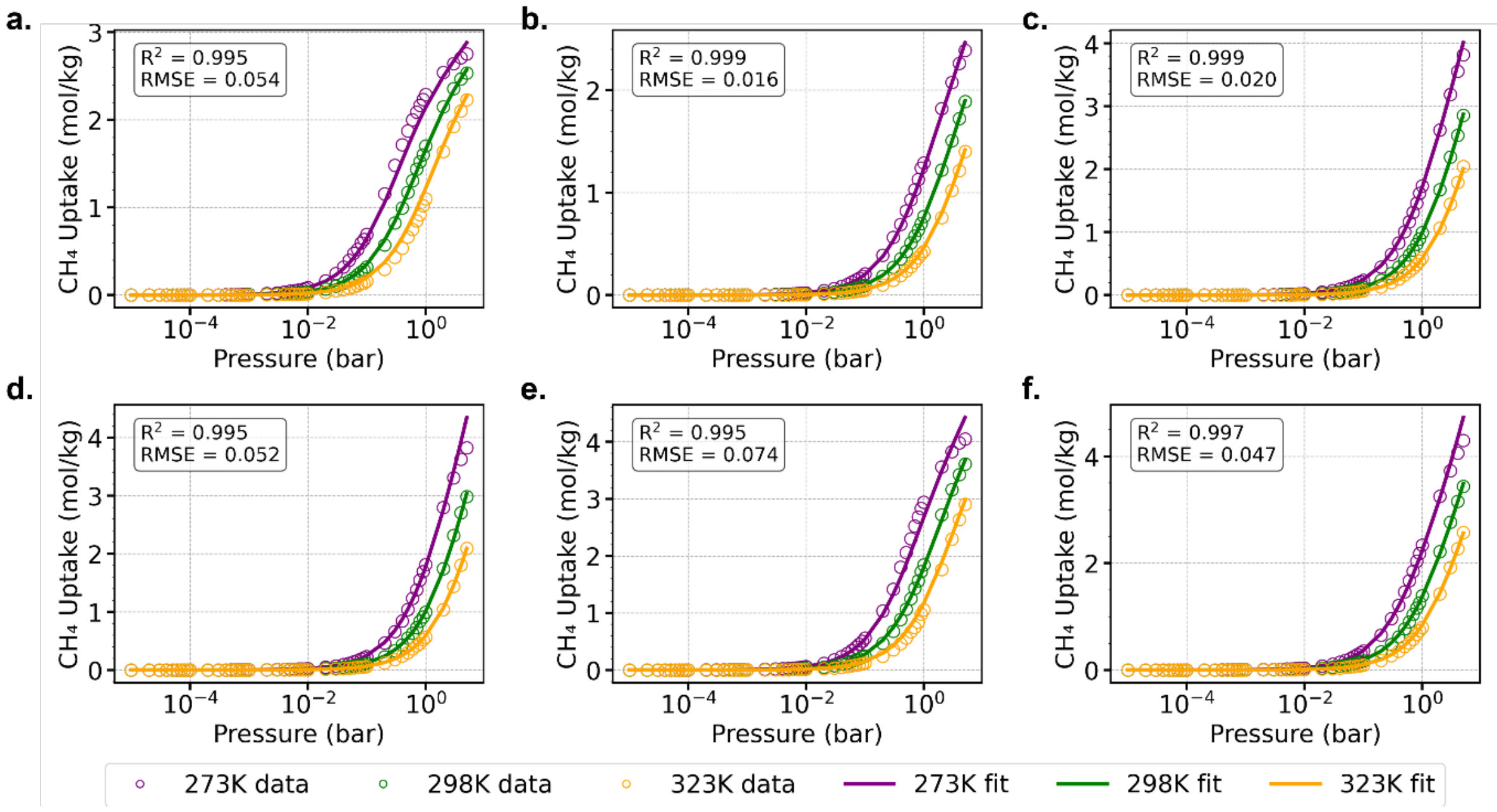

**Fig. S4** Curve fitting results of $CH_4$ isotherms for **a.** CALF-20, **b.** SquCALF-20, **c.** FumCALF-20, **d.** Bdc-CALF-20, **e.** CubCALF-20, and **f.** Ttdc-CALF-20 using the dual-site Langmuir (DSL) model at 273 K, 298 K, and 323 K. Solid lines represent DSL fits, and open symbols denote the corresponding data points.

## Section S8. DSL isotherm parameters for $CO_2$ and $CH_4$

**Table S1.** DSL isotherm parameters from curve fitting.

| Materials | Parameter | $CO_2$ | $CH_4$ |
|---|---|---|---|
| CALF-20 | $q_b$ (mol/kg) | 3.08 | 3.08 |
| | $q_d$ (mol/kg) | 2.77 | 2.77 |
| | $b_0$ ($m^3$/mol) | $3.41 \times 10^{-11}$ | $3.02 \times 10^{-11}$ |
| | $d_0$ ($m^3$/mol) | $3.59 \times 10^{-9}$ | $4.40 \times 10^{-9}$ |
| | $\Delta U_b$ (J/mol) | −30,094 | −20,013 |
| | $\Delta U_d$ (J/mol) | −29,093 | −20,039 |
| SquCALF-20 | $q_b$ (mol/kg) | 2.55 | 2.55 |
| | $q_d$ (mol/kg) | 3.01 | 3.01 |
| | $b_0$ ($m^3$/mol) | $7.28 \times 10^{-10}$ | $1.20 \times 10^{-9}$ |
| | $d_0$ ($m^3$/mol) | $3.56 \times 10^{-10}$ | $4.84 \times 10^{-11}$ |
| | $\Delta U_b$ (J/mol) | −25,010 | −20,013 |
| | $\Delta U_d$ (J/mol) | −24,019 | −20,000 |
| FumCALF-20 | $q_b$ (mol/kg) | 4.34 | 4.34 |
| | $q_d$ (mol/kg) | 3.43 | 3.43 |
| | $b_0$ ($m^3$/mol) | $1.16 \times 10^{-10}$ | $1.43 \times 10^{-10}$ |
| | $d_0$ ($m^3$/mol) | $1.74 \times 10^{-10}$ | $9.43 \times 10^{-10}$ |
| | $\Delta U_b$ (J/mol) | −30,000 | −19,997 |
| | $\Delta U_d$ (J/mol) | −29,000 | −20,007 |
| BdcCALF-20 | $q_b$ (mol/kg) | 6.77 | 6.77 |
| | $q_d$ (mol/kg) | 4.11 | 4.11 |
| | $b_0$ ($m^3$/mol) | $6.56 \times 10^{-11}$ | $7.29 \times 10^{-11}$ |
| | $d_0$ ($m^3$/mol) | $9.68 \times 10^{-12}$ | $8.13 \times 10^{-10}$ |
| | $\Delta U_b$ (J/mol) | −30,028 | −20,008 |
| | $\Delta U_d$ (J/mol) | −29,005 | −20,016 |
| CubCALF-20 | $q_b$ (mol/kg) | 5.01 | 5.01 |
| | $q_d$ (mol/kg) | 4.46 | 4.46 |
| | $b_0$ ($m^3$/mol) | $2.47 \times 10^{-10}$ | $3.61 \times 10^{-11}$ |
| | $d_0$ ($m^3$/mol) | $3.32 \times 10^{-12}$ | $1.99 \times 10^{-9}$ |
| | $\Delta U_b$ (J/mol) | −30,025 | −20,003 |
| | $\Delta U_d$ (J/mol) | −29,000 | −20,008 |
| TtdcCALF-20 | $q_b$ (mol/kg) | 8.35 | 8.36 |
| | $q_d$ (mol/kg) | 3.91 | 3.91 |
| | $b_0$ ($m^3$/mol) | $8.62 \times 10^{-11}$ | $6.59 \times 10^{-11}$ |
| | $d_0$ ($m^3$/mol) | $7.13 \times 10^{-12}$ | $1.36 \times 10^{-9}$ |
| | $\Delta U_b$ (J/mol) | −30,002 | −19,998 |
| | $\Delta U_d$ (J/mol) | −28,999 | −20,024 |

## Section S9. Binary mixture adsorption isotherms

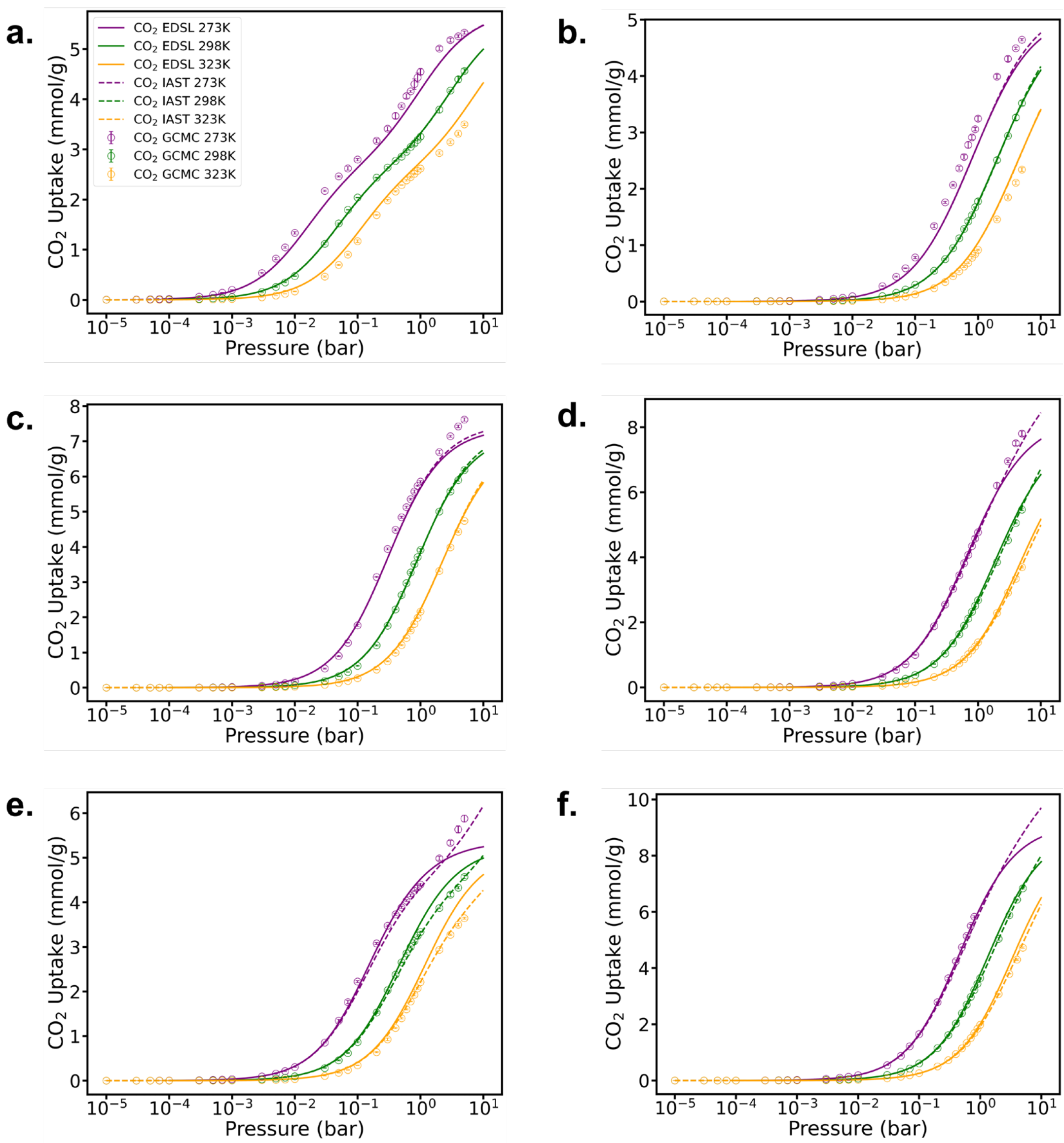

**Fig. S5** Comparison of $CO_2$ binary mixture ($CH_4$:$CO_2$ = 50:50) adsorption isotherms predicted by the extended dual-site Langmuir (EDSL) model and the ideal adsorbed solution theory (IAST), together with GCMC simulation results at 273 K, 298 K, and 323 K for six CALF-20 derivatives: (a) CALF-20, (b) SquCALF-20, (c) FumCALF-20, (d) BdcCALF-20, (e) CubCALF-20, and (f) TtdcCALF-20.

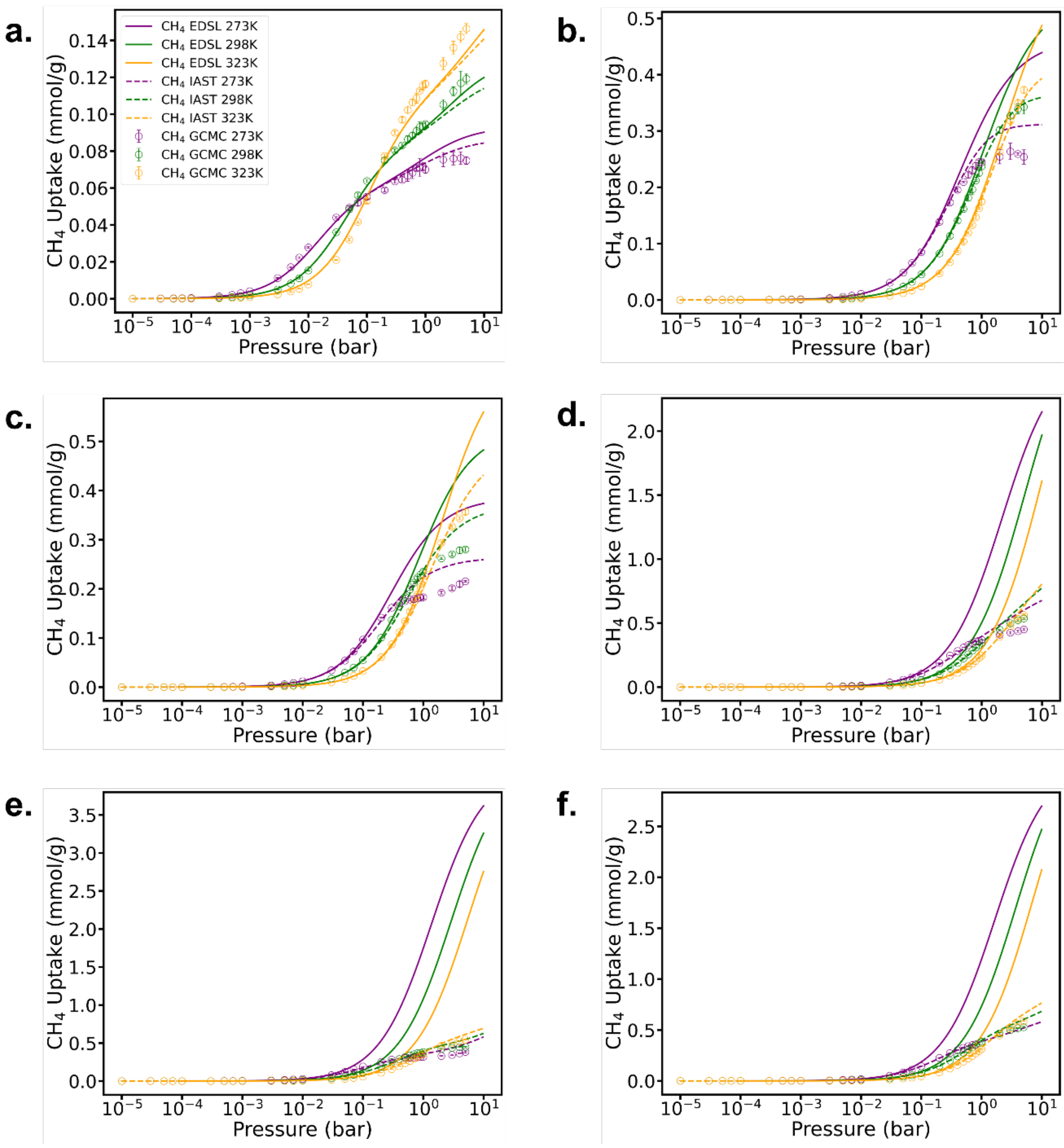

**Fig. S6** Comparison of $CH_4$ binary mixture ($CH_4$:$CO_2$ = 50:50) adsorption isotherms predicted by the extended dual-site Langmuir (EDSL) model and the ideal adsorbed solution theory (IAST), together with GCMC simulation results at 273 K, 298 K, and 323 K for six CALF-20 derivatives: (a) CALF-20, (b) SquCALF-20, (c) FumCALF-20, (d) BdcCALF-20, (e) CubCALF-20, and (f) TtdcCALF-20.

## Section S10. Literature data for benchmark $CO_2/CH_4$ separation adsorbents

**Table S2.** Summary of literature-reported properties and separation performance of benchmark adsorbents for $CO_2/CH_4$ mixtures.

| Adsorbent | Surface area ($m^2/g$) | Pore volume ($cm^3/g$) | Process condition | | | Selectivity ($CO_2/CH_4$) | References |
|---|---|---|---|---|---|---|---|
| | | | Temperature (K) | Pressure (bar) | $y_{CH_4}$ | | |
| MOF-5 | 3,054 | 1.31 | 297 | 15 | 0.50 | 3.06 | Kloutse et al.[3] |
| | | | 297 | 10 | 0.50 | 2.61 | |
| | | | 297 | 5 | 0.50 | 2.49 | |
| | | | 297 | 15 | 0.20 | 2.93 | |
| | | | 297 | 15 | 0.80 | 2.50 | |
| | | | 297 | 5 | 0.20 | 2.44 | |
| | | | 297 | 5 | 0.80 | 2.53 | |
| Zeolite 13X | 534 | 0.29 | 277 | 1 | 0.60 | 8.47 | Pour et al.[4] |
| Zeolite NaX | 534 | 0.29 | 303 | 10 | 0.50 | 76 | Yundong et al.[5] |
| Zeolite CaA | 397 | 0.32 | 303 | 10 | 90 | 74 | |
| Zeolite 5A | 457–600 | – | 303 | 1.2 | 0.60 | 17.25 | Mofarahi et al.[6] |
| | | | | 4.4 | 0.60 | 13.50 | |
| | | | 323 | 1.2 | 0.60 | 8.50 | |
| | | | | 4.4 | 0.60 | 19.94 | |
| Mesophase pitches-based CMSs | | | | | | | |
| VR-5-P | 3,100 | 1.57 | 298 | 1 | – | 7.0 | Wahby et al.[7] |
| VR-5-M | 2,450 | 1.12 | 298 | 1 | – | 4.1 | |
| VR-93-P | 2,720 | 1.22 | 298 | 1 | – | ∞ | |
| VR-93-M | 2,895 | 1.42 | 298 | 1 | – | ∞ | |
| DO-10-mix | 2,595 | 1.16 | 298 | 1 | – | 7.4 | |
| DO-88-P | 2,445 | 1.14 | 298 | 1 | – | 6.6 | |
| DO-88-M | 2,660 | 1.16 | 298 | 1 | – | 6.2 | |
| CMSs from palm shell | | | | | | | |
| AC250 | 1,104 | 0.41 | 298 | 1 | – | 3.27 | Ahmad et al.[8] |
| CMS600 | 932 | 0.38 | 298 | 1 | – | 6.66 | |
| CMS700 | 847 | 0.39 | 298 | 1 | – | 11.80 | |
| CMS800 | 704 | 0.41 | 298 | 1 | – | 16.00 | |
| CMS900 | 621 | 0.29 | 298 | 1 | – | 11.55 | |

**Section S11. Ternary adsorption isotherms from GCMC simulations**

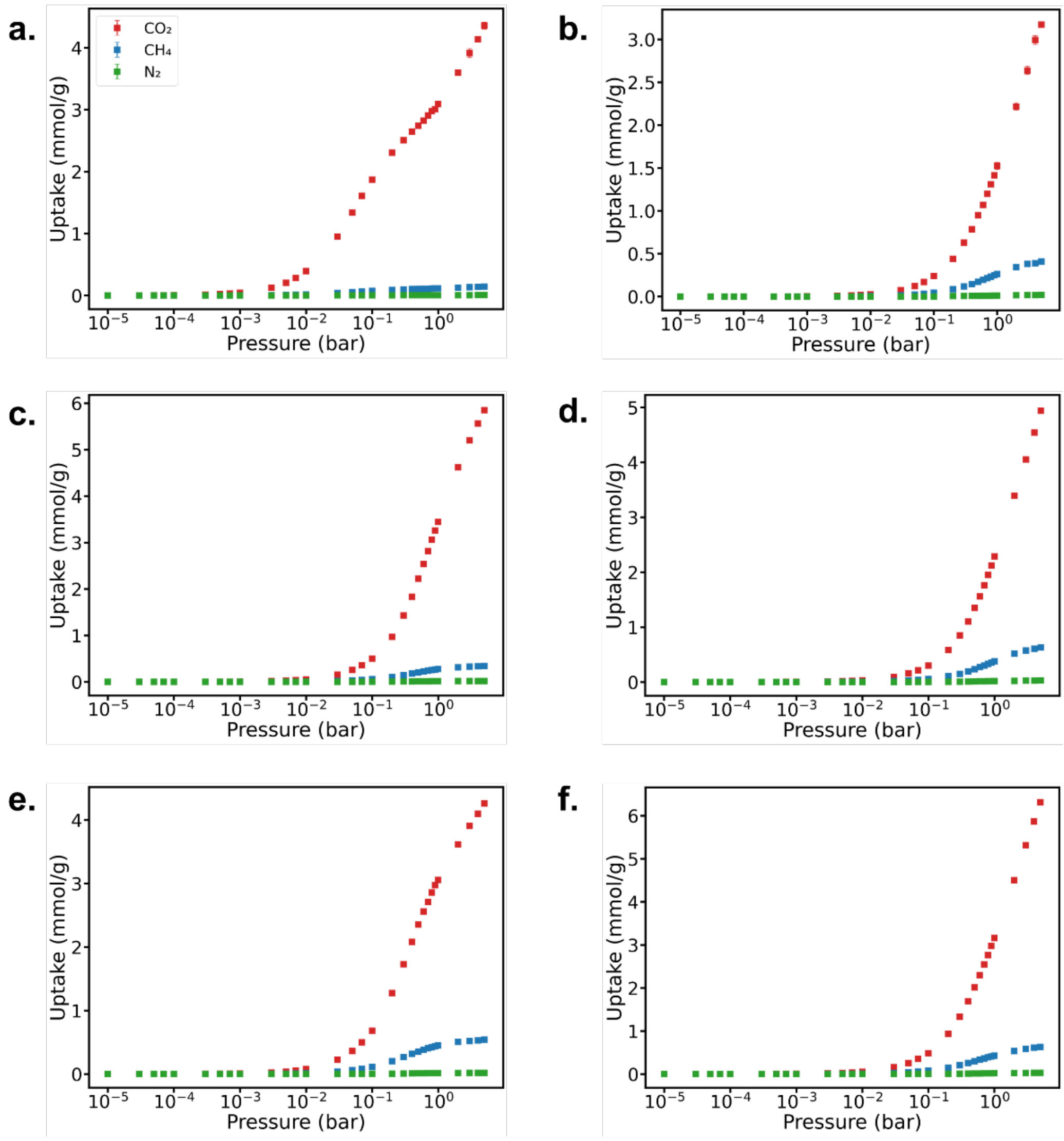

**Fig. S7** Ternary adsorption isotherms of $CO_2$ (red), $CH_4$ (blue), and $N_2$ (green) at 298 K obtained from ternary GCMC simulations for six CALF-20 derivatives: (a) CALF-20, (b) SquCALF-20, (c) FumCALF-20, (d) BdcCALF-20, (e) CubCALF-20, and (f) TtdcCALF-20.

## Section S12. Additional techno-economic details of the PVSA process

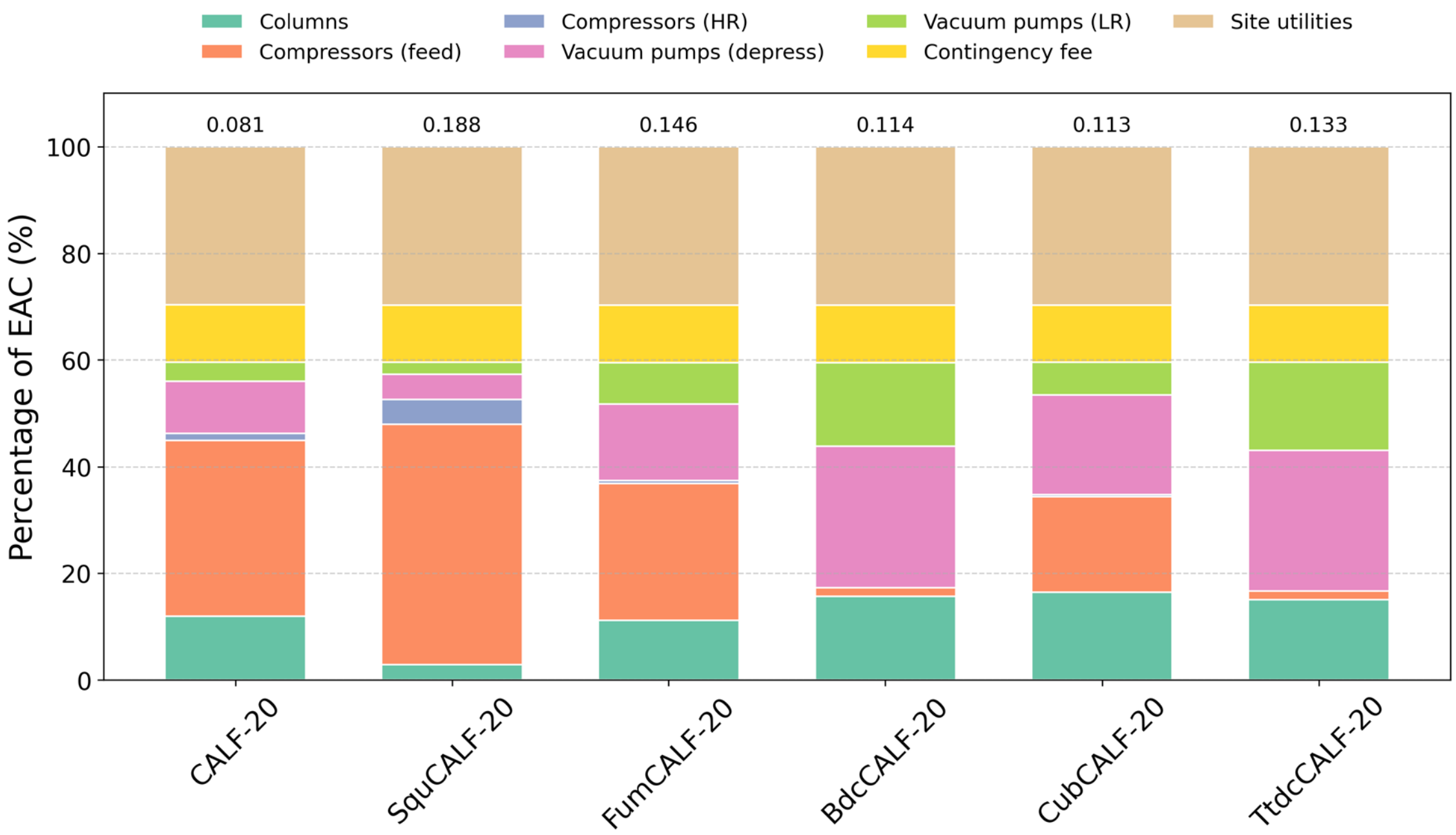

**Fig. S8** Percentage contributions of EAC to the methane production cost of six CALF-20 derivatives under cost-optimal PVSA operation. The absolute EAC values (in \$/kg $CH_4$) are indicated above each bar.

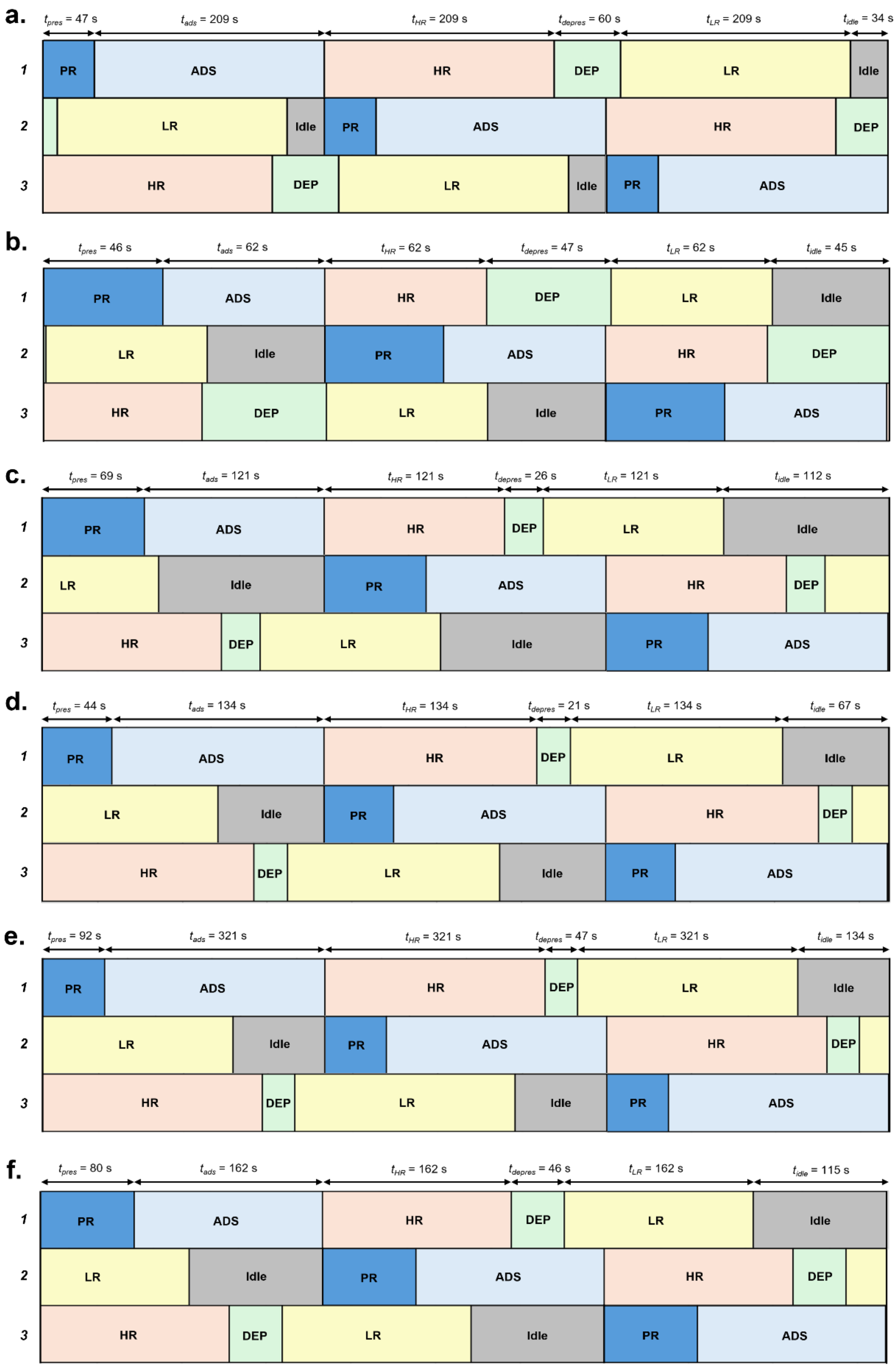

**Fig. S9** Cyclic scheduling diagram of the three-column PVSA unit train under the optimized five-step modified Skarstrom cycle for a. CALF-20, b. SquCALF-20, c. FumCALF-20, d. BdcCALF-20, e. CubCALF-20, and f. TtdcCALF-20.

**Section S13. Molecular-level adsorption analysis**

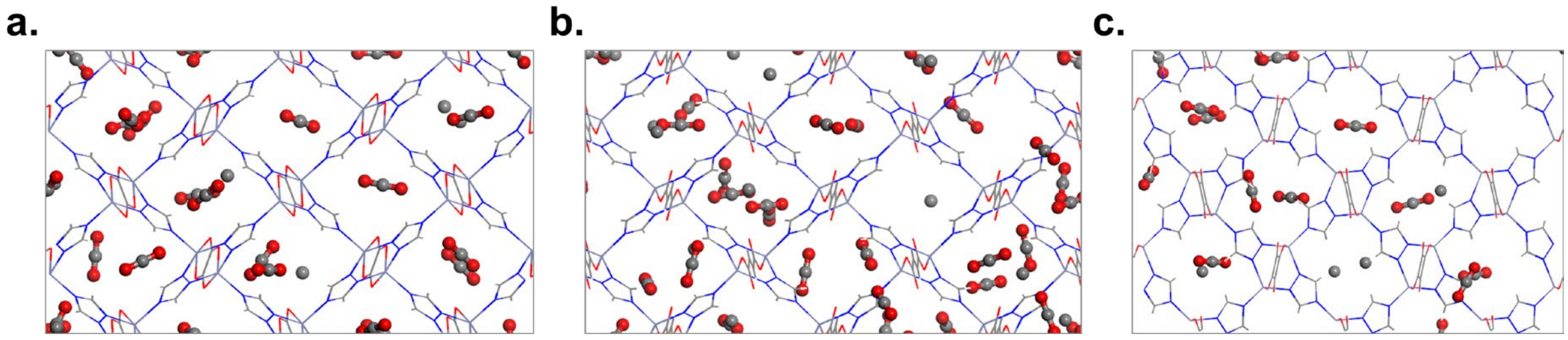

**Fig. S10** Representative molecular simulation snapshots of $CO_2$ and $CH_4$ distributions in: a. CALF-20, b. FumCALF-20, and c. SquCALF-20 at 298 K and 1 bar. Atom colors: gray (C) and red (O). Guest molecules: linear O–C–O corresponds to $CO_2$, while single gray spheres correspond to $CH_4$.

Fig. S10 shows the distributions of adsorbed $CO_2$ and $CH_4$ in CALF-20, FumCALF-20, and SquCALF-20. In CALF-20, the pores are predominantly filled with $CO_2$ while $CH_4$ is only rarely observed, indicating a pronounced selective adsorption behavior arising from the narrow pore structure and strong interactions between $CO_2$ and the framework. In FumCALF-20, $CO_2$ also dominates the adsorption, and snapshots from another viewing angle reveal that multiple $CO_2$ molecules can coexist within a single pore, reflecting the larger pore volume and higher $CO_2$ capacity of this material. In contrast, SquCALF-20 exhibits squarate linkers that are not aligned parallel to the pore axis but are slightly tilted, which makes it more difficult for molecules to access and efficiently occupy the pore space. Such an arrangement may induce steric effects during molecular migration, which can be regarded as one of the structural factors associated with the lower separation performance of SquCALF-20 compared to CALF-20 and FumCALF-20.

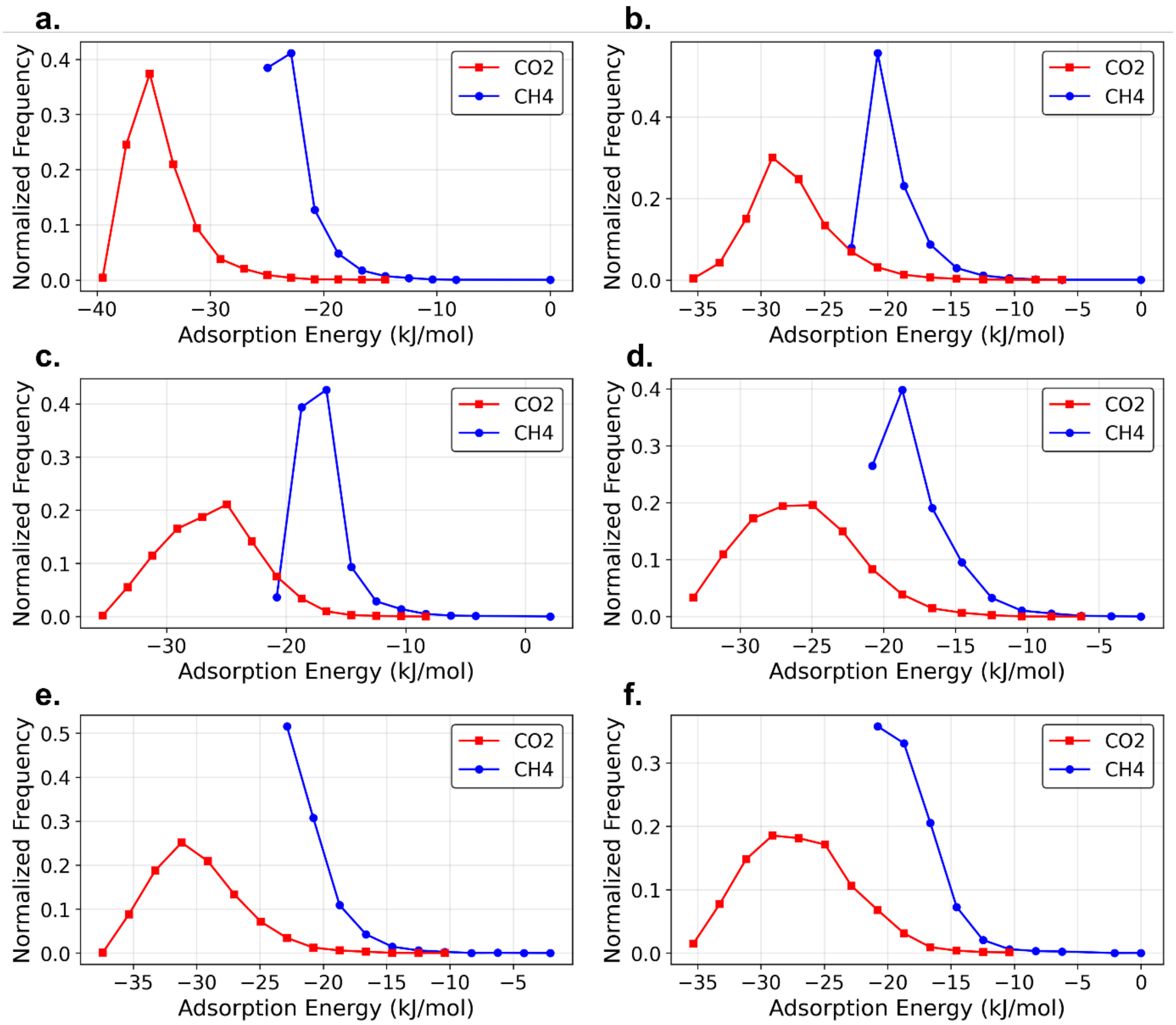

**Fig. S11** Adsorption energy distributions of $CO_2$ (red) and $CH_4$ (blue) obtained from GCMC simulations at 298 K for six CALF-20 derivatives: (a) CALF-20, (b) SquCALF-20, (c) FumCALF-20, (d) BdcCALF-20, (e) CubCALF-20, and (f) TtdcCALF-20.

The adsorption energy distribution analysis revealed that in all CALF-20 derivatives, $CO_2$ is centered around relatively deep energy states near –30 kJ/mol, whereas $CH_4$ is confined to shallower states around –20 kJ/mol (Fig. S11). This indicates that $CO_2$ forms stronger interactions with the framework and preferentially occupies stable adsorption sites, while $CH_4$ remains only weakly stabilized and more limited in distribution. In particular, CALF-20 shows a larger gap between the $CO_2$ and $CH_4$ peaks than the other derivatives, which can be interpreted as providing more favorable conditions for selective adsorption.

Meanwhile, in BdcCALF-20, CubCALF-20, and TtdcCALF-20, the gap between the $CO_2$ and $CH_4$ peaks is also relatively large; however, these materials exhibit flexible linker conformations or elongated pore channels that weaken the confinement effect. As a result, even if $CO_2$ preferentially occupies stable sites, $CH_4$ can still coexist within the pores, leading to lower overall selectivity. Therefore, the high selectivity observed in CALF-20 can be understood not only as a consequence of the energy distribution difference, but also as a combined effect of the narrow and confined pore structure suppressing competitive $CH_4$ adsorption.

## References

(1) Zanco, S. E.; Pérez-Calvo, J.-F.; Gasós, A.; Cordiano, B.; Becattini, V.; Mazzotti, M. Postcombustion CO2 capture: a comparative techno-economic assessment of three technologies using a solvent, an adsorbent, and a membrane. *ACS Engineering Au* **2021**, *1* (1), 50-72. DOI: 10.1021/acsengineeringau.1c00002.

(2) Ward, A.; Pini, R. Efficient bayesian optimization of industrial-scale pressure-vacuum swing adsorption processes for co2 capture. *Industrial & Engineering Chemistry Research* **2022**, *61* (36), 13650-13668. DOI: 10.1021/acs.iecr.2c02313.

(3) Kloutse, F.; Hourri, A.; Natarajan, S.; Benard, P.; Chahine, R. Experimental benchmark data of CH4, CO2 and N2 binary and ternary mixtures adsorption on MOF-5. *Separation and Purification Technology* **2018**, *197*, 228-236. DOI: 10.1016/j.seppur.2018.01.013.

(4) Pour, A. A.; Sharifnia, S.; NeishaboriSalehi, R.; Ghodrati, M. Performance evaluation of clinoptilolite and 13X zeolites in CO2 separation from CO2/CH4 mixture. *Journal of Natural Gas Science and Engineering* **2015**, *26*, 1246-1253. DOI: 10.1016/j.jngse.2015.08.03.

(5) Li, Y.; Yi, H.; Tang, X.; Li, F.; Yuan, Q. Adsorption separation of CO2/CH4 gas mixture on the commercial zeolites at atmospheric pressure. *Chemical Engineering Journal* **2013**, *229*, 50-56. DOI: 10.1016/j.cej.2013.05.101.

(6) Mofarahi, M.; Gholipour, F. Gas adsorption separation of CO2/CH4 system using zeolite 5A. *Microporous and Mesoporous Materials* **2014**, *200*, 1-10. DOI: 10.1016/j.micromeso.2014.08.022.

(7) Wahby, A.; Ramos-Fernández, J. M.; Martínez-Escandell, M.; Sepúlveda-Escribano, A.; Silvestre-Albero, J.; Rodríguez-Reinoso, F. High-surface-area carbon molecular sieves for selective CO2 adsorption. *ChemSusChem* **2010**, *3* (8), 974-981. DOI: 10.1002/cssc.201000083.

(8) Ahmad, M. Preparation of carbon molecular sieves from palm shell: effect of benzene deposition conditions. *Adsorption* **2009**, *15* (5), 489-495. DOI: 10.1007/s10450-009-9199-0.